\documentclass[a4paper, 11pt, oneside]{article}
\usepackage{float}
\usepackage{graphicx}
\usepackage{amsmath,amssymb,epsfig,graphicx}
\usepackage{epstopdf}
\usepackage{jheppub}
\newcommand{\nc}{\newcommand}
\nc{\non}{\nonumber}
\nc{\hc}{\hbox {H.c.}}
\nc{\noi}{\noindent}
\nc{\barx}{\bar{x}}
\nc{\pbarn}{\;\hbox {pb}}
\nc{\fbarn}{\;\hbox {fb}}
\nc{\lsp}{\;\;\;\;\;}
\nc{\Lsp}{\;\;\;\;\;\;\;\;\;\;}
\nc{\LLsp}{\lspace \lspace}
\nc{\lra}{\longrightarrow}
\nc{\beq}{\begin{equation}}  \nc{\eeq}{\end{equation}}
\nc{\bea}{\begin{eqnarray}}  \nc{\eea}{\end{eqnarray}}
\nc{\baa}{\begin{array}}     \nc{\eaa}{\end{array}}
\nc{\bit}{\begin{itemize}}   \nc{\eit}{\end{itemize}}
\nc{\ben}{\begin{enumerate}} \nc{\een}{\end{enumerate}}
\nc{\bce}{\begin{center}}    \nc{\ece}{\end{center}}
\nc{\bpm}{\begin{pmatrix}}   \nc{\epm}{\end{pmatrix}}
\nc{\bvt}{\begin{verbatim}}  \nc{\evt}{\end{verbatim}}

\def\lsim{\mathrel{\raise.3ex\hbox{$<$\kern-.75em\lower1ex\hbox{$\sim$}}}}
\def\gsim{\mathrel{\raise.3ex\hbox{$>$\kern-.75em\lower1ex\hbox{$\sim$}}}}

\def\udots{\mathinner{\mkern1mu\raise1pt\vbox{\kern7pt\hbox{.}}\mkern2mu\raise4pt\hbox{.}\mkern2mu\raise7pt\hbox{.}\mkern1mu}}

\def\tev{\;\hbox{TeV}}

\newcommand\fverb{\setbox\fverbbox=\hbox\bgroup\verb}
\newcommand\fverbdo{\egroup\medskip\noindent%
			\fbox{\unhbox\fverbbox}\ }
\newcommand\fverbit{\egroup\item[\fbox{\unhbox\fverbbox}]}
\newbox\fverbbox

\title{Brane modeling in warped extra-dimension}

\author[\ast]{Aqeel Ahmed\note[$\ast$]{On leave of absence from National Centre for Physics,
Quaid-i-Azam University Campus, Islamabad 45320, Pakistan}}
\author[]{and Bohdan Grzadkowski}
\affiliation[]{Faculty of Physics,
University of Warsaw,\\
Ho\.za 69, 00-681 Warsaw, Poland}
\emailAdd{aqeel.ahmed@fuw.edu.pl}
\emailAdd{bohdan.grzadkowski@fuw.edu.pl}
\date{\today}

\abstract{Five-dimensional scenarios with infinitesimally thin branes replaced by appropriate configurations
of a scalar field were considered. A possibility of periodic extra dimension was discussed
in the presence on non-minimal scalar-gravity coupling and a generalized
Gibbons-Kallosh-Linde sum rule was found.
In order to avoid constraints imposed by periodicity, a non-compact spacial extra dimension was introduced.
A five dimensional model with warped geometry and two thin branes mimicked by a scalar profile
was constructed and discussed. In the thin brane limit the model corresponds to a set-up with two
positive-tension branes. The presence of two
branes allows to address the issue of the hierarchy problem which could be solved by the standard
warping of the four dimensional metric provided the Higgs field is properly localized.
Stability of the background solution was discussed
and verified in the presence of the most general perturbations of the metric and the scalar field.
}

\arxivnumber{1210.6708}

\begin{document}
\maketitle
\flushbottom


\section{Introduction}
\label{Introduction}

The idea of extra-dimensions of space-time inspired by the string theory received a lot of attention since
the last decade or so as a possible solution to the problem of hierarchy between the Planck scale
$M_{Pl}\approx10^{18}$ GeV and the electroweak scale
$m_{EW}\approx10^3$ GeV \cite{ArkaniHamed:1998rs,Antoniadis:1998ig,Randall:1999ee}. The most attractive incarnation
of the idea, proposed by Randall and
Sundrum \cite{Randall:1999ee}, involves one extra-dimension with non-trivial warped factor
appearing due to the anti-de Sitter (AdS) geometry along the fifth-dimension. This is so-called RS1,
the Randall-Sundrum model with two D3 branes on the $S_1/Z_2$ orbifold along the extra-dimension.
In this model the hierarchy problem can be addressed without introducing the large compactified
volume of the extra-dimensions as suggested by Arkani-Hamed, Dimopoulos, Dvali (ADD) \cite{ArkaniHamed:1998rs,Antoniadis:1998ig}, by the virtue of the non-trivial warped geometry
along the extra-dimension. The model requires the presence of two singular D3 branes of opposite
tension. There were many attempts to avoid the presence of thin (singular) branes. It has been shown~\cite{Rubakov:1983bb,DeWolfe:1999cp,Gremm:1999pj,Csaki:2000fc,Kehagias:2000au,Kobayashi:2001jd,Bronnikov:2003gg,Bazeia:2008zx,Bazeia:2007nd,Melfo:2002wd,BarbosaCendejas:2007hs,Dzhunushaliev:2009va},
that the positive tension brane could be smoothed (or, in other words,
regularized) by a background scalar field configuration which we term here as the smooth or thick brane.
However there is no satisfactory simple strategy to model the negative tension brane at least for
scalar field minimally coupled to gravity. For example of an existing attempts to generate negative tension branes see \cite{Oda:2008tk}. Also, as shown by Gibbons et al. \cite{Gibbons:2000tf},
periodicity of set-ups like RS1 is generically in conflict with the idea of a smooth non-trivial
scalar profile. So, concluding, smoothing the RS1 scenario is severely limited by periodicity
and impossibility of generating a negative tension brane by a scalar field configuration. The conclusion holds
at least for the case when the scalar field is minimally coupled to gravity, the generalization will be considered
below.

It has been pointed out by Randall and Sundrum in their second seminal paper that the extra-dimension
can be infinite and yet it can lead to nearly standard 4D gravity~\cite{Randall:1999vf}.
The main idea in this second paper (RS2) is that a single D3 brane of {\it positive tension}
is embedded in a 5D AdS geometry and the gravity is effectively 4D at large
distances. The 4D graviton is localized on the brane which implies finite 4D Planck mass.
Since a positive tension brane could be mimicked by a scalar field, therefore the RS2 seems to be
more attractive from that perspective. However, the RS2 model having just one D3 brane
is not addressing the hierarchy problem. Since solving the hierarchy problem is
the main motivation for our study, therefore we will try to improve the scenario
by introducing a second brane.

The purpose of present work is twofold:
\bit
\item to see if one can overcome the above mentioned obstacles (periodicity and positivity of brane tension)
to achieve a smooth version of RS1 in modified gravity with the scalar field non-minimally
coupled to the Ricci scalar, and
\item to verify if one can address the hierarchy problem with two thick branes (which in a certain
``brane limit'' mimic two positive tension singular branes) with non-compact warped
extra-dimension.
\eit
As far as the first question is concerned, we will show, by generalizing the Gibbons-Kallosh-Linde
sum rules \cite{Gibbons:2000tf}, that it is not possible to achieve the periodicity
for scalar field and metric solutions even with the scalar field non-minimally coupled to
the Ricci scalar. The consistency conditions have been discussed
following the strategy of Gibbons et al. \cite{Gibbons:2000tf} in the
modified gravity set-up \cite{Abdalla:2010sz}, however the authors consider the scalar field
in the bulk with {\it singular branes}, this is exactly what we want to avoid for our set-up, i.e., we
wish to have smooth branes instead of singular branes. Another attempt to overcome the
problem of periodicity was discussed in \cite{Wudka:2011we}.
Concerning the issue of the positivity of a brane tension generated by a scalar profile
we also find that even with non-minimal scalar couplings there is no way to generate a
negative tension brane. Therefore we turn our attention to models with only positive tension branes
(e.g. the RS2) and non-compact (to avoid restrictions imposed by periodicity a'la \cite{Gibbons:2000tf}).
Since we find a satisfactory model with a scalar that is minimally coupled to gravity,
we restrict ourself to that scenario. In order to be able to address the hierarchy problem
we will propose a model with two thick (smooth) branes, which, in an appropriately defined limit
(so called brane limit) approaches two singular branes. The limiting version of the model was
discussed earlier by Lykken and Randall (LR) in \cite{Lykken:1999nb}.
As we will show, in our set-up of two thick branes, different possible solutions for the warped factor
can emerge, for instance, we can have the AdS or
Minkowski geometry in different regions along the extra-dimension. We will discuss three
such configurations, {\it (i)} the two thick branes between the AdS vacua so that we have warped
geometry and hierarchy problem could be addressed in this set-up (this is the thick brane
version of the Lykken-Randall model \cite{Lykken:1999nb}), {\it (ii)} the case when we can have
the Minkowski background in between the two branes and the AdS geometry to the right and
left of both branes and, {\it (iii)} with the Minkowski geometry to right or left of both branes
and the AdS in the other regions along the extra-dimension, which gives the thick brane
version of the Gregory, Rubakov, Sibiryakov (GRS) model \cite{Gregory:2000jc} except that we
have the second brane also with positive tension instead of the negative tension as in
the original GRS model with singular branes.
In all the above mentioned three cases we study stability of the background solutions and also existence and
localization of the zero-modes of the scalar, vector and tensor (SVT) perturbations of
the solutions. The issue of stability and localization of the zero-modes have been extensively discussed in the past in the context of thin as well as thick brane scenarios \cite{DeWolfe:1999cp,Gremm:1999pj,Csaki:2000fc,Kehagias:2000au,Kobayashi:2001jd,Garriga:1999yh,Giddings:2000mu,Karch:2000ct,Csaki:2000ei,DeWolfe:2000xi,Giovannini:2001fh,Giovannini:2001xg, Csaki:2000zn,Cvetic:2008gu,Kakushadze:2000zp,Brandhuber:1999hb,Aybat:2010sn,Andrianov:2012ae}.
We find that the SVT perturbation equations could be transformed into a supersymmetric quantum
mechanics form so that they guarantee stability of these perturbations in all the
configurations considered above and also the absence of tachyonic modes. It turns
out that the zero-mode of the tensor perturbation wave function
(that corresponds to the 4D gravitons) is localized in the cases {\it (i)} and {\it (ii)}
but it is quasi-localized in the case {\it (iii)} mentioned above. The zero-modes
corresponding to the scalar and vector perturbations are not localized, as they are not
normalizable modes, consequently they do not affect the 4D physics.

The paper is organized as follows. Possibilities and difficulties
of constructing smooth generalizations of the RS1 in the modified gravity scenario
is considered in Sec~\ref{Thick
brane generalization of RS1 in modified gravity}. In Sec.~\ref{Two thick branes and the
background solutions} the background solutions of the two thick brane set-up is discussed
in the standard Einstein-Hilbert gravity. Possible different geometric configurations
for the two thick branes are discussed in Sec.~\ref{The brane limit} along with their brane limit solutions.
Sec.~\ref{Linearized Einstein equations} is dedicated to the linearized Einstein
equations. Where we have
derived the equations of motion for the SVT perturbations in the Newtonian or longitudinal gauge. The issues of stability and
localization of zero-modes of
the SVT perturbations are covered in Sec.~\ref{Stability of the SVT perturbations}. The
last section, Sec.~\ref{Conclusions}, contains our conclusions.
Appendices are dedicated to conventions and some mathematical details for the SVT decomposition
of the metric perturbations and the gauge choice.


\section{Thick brane generalization of RS1 in modified gravity}
\label{Thick brane generalization of RS1 in modified gravity}

Our first goal is to mimic (regularize) D3-branes which appear in various five-dimensional (5D) scenarios
that solve the hierarchy problem by warping the metric along the extra dimension in the spirit of
\cite{Randall:1999ee}. The most natural approach is to introduce a 5D scalar field $\phi$ with
a non-trivial profile (that satisfies equations of motion) that in certain limit could
mimic a brane by approaching a delta-like energy distribution along extra dimension.
However, as it was shown in \cite{Gibbons:2000tf}, in the case of compact extra dimensions
the idea of a non-trivial scalar profile (a thick brane) is severely restricted by the requirement of periodicity.
Arguments adopted in \cite{Gibbons:2000tf} apply for a scalar that is minimally coupled to gravity.
Therefore here, we are going to discuss first a class of models allowing for non-minimal scalar-gravity
coupling:
\begin{equation}
S=\int dx^5 \sqrt{-g}\left\{f(\phi)R-\frac{1}{2}g^{MN}\nabla_{M}\phi\nabla_{N}\phi-V(\phi)\right\},
\label{action_m}
\end{equation}
where $f(\phi)$ is a general smooth positive definite function of the scalar field
$\phi$\footnote{From the action \eqref{action_m} one can infer that $\phi(y)$
and $V(\phi)$ have the mass dimensions 3/2 and 5, respectively and also we assume that the scalar field $\phi$ has only dependence on the extra-spacial coordinate $y$.},
which is supposed to compose the D3-branes that are present in the RS1 scenario
\cite{Randall:1999ee}. In our convention capital roman indices will refer to 5D objects,
i.e., $M,N,\cdots=0,1,2,3,5$ whereas,
the Greek indices label four-dimensional (4D) objects, i.e., $\mu,\nu,\cdots=0,1,2,3$.
In the Eq.~\eqref{action_m} $\nabla_{M}$
is 5D covariant derivative, for conventions and details see the Appendix \ref{Conv} .

In other words, branes would be made of the scalar field while other fields could be dynamically
localized in certain regions of the 5D space, see for instance
\cite{Davoudiasl:1999tf,Pomarol:1999ad,Grossman:1999ra,Chang:1999nh}.
Thick branes in the presence of non-minimally coupled scalar was discussed earlier by \cite{Bogdanos:2006qw,Guo:2011wr}.

We will look for a solution of the Einstein equations with the following 5D
metric\footnote{We choose the metric signature $(- + + + +)$.} ansatz
\begin{equation}
ds^2=e^{2A(y)}\eta_{\mu\nu}dx^{\mu} dx^{\nu}+dy^2, \label{metric}
\end{equation}
where $\eta_{\mu\nu}(x)$ is the 4D metric and the warp function $A(y)$ is only a function of the extra-spatial coordinate $y$.
The Einstein's equations of motion, resulting from the action \eqref{action_m} are
\begin{eqnarray}
R_{MN}-\frac{1}{2}g_{MN}R&=&\frac{1}{f(\phi)}\left\{\frac{1}{2}T_{MN}+\nabla_{M}\nabla_{N}f(\phi)
-g_{MN}\nabla^{2}f(\phi)\right\},\label{eineq1_m}\\
\nabla^{2}\phi-\frac{dV}{d\phi}+R\frac{df}{d\phi}&=&0, \label{eineq2_m}
\end{eqnarray}
with the energy-momentum tensor for the scalar field $\phi$ as
\begin{equation}
T_{MN}=\nabla_{M}\phi\nabla_{N}\phi-g_{MN}\left(\frac{1}{2}(\nabla\phi)^{2}+V(\phi)\right).\label{emt_m}
\end{equation}
From the Einstein equations \eqref{eineq1_m} and \eqref{eineq2_m}, one can get the equations of motion for
the metric ansatz \eqref{metric} as,
\begin{align}
6(A^{\prime})^{2}&=\frac{1}{f}\left\{\frac{1}{4}(\phi^{\prime})^{2}-\frac{1}{2}V
-4A^{\prime}f^{\prime}\right\},\label{eom01_m}\\
3A^{\prime\prime}+6(A^{\prime})^{2}&=\frac{1}{f}\left\{-\frac{1}{4}(\phi^{\prime})^{2}-\frac{1}{2}V
-3A^{\prime}f^{\prime}-f^{\prime\prime}\right\},\label{eom02_m}\\
\phi^{\prime\prime}+4A^{\prime}\phi^{\prime}&-\frac{dV}{d\phi}-\left(8A^{\prime\prime}+20(A^{\prime})^{2}\right)\frac{df}{d\phi}=0, \label{eom03_m}
\end{align}
where prime $^\prime$ denoted the derivative with respect to $y$, and it is understood that $f$ and $V$ are functions of the scalar field $\phi(y)$.

\subsection{Thick branes with periodic extra dimensions}
\label{Thick branes with periodic extra dimensions}

Here we would like to verify if the scenario with a non-trivial profile of a bulk
scalar could be consistent with periodicity in the case of compact extra dimension.
The authors of \cite{Gibbons:2000tf} derived elegant, simple and powerful sum rules that severely
restrict thick brane scenarios with periodic extra dimensions. From our perspective the most relevant result obtained there
is the following condition that must be satisfied for periodic extra dimensions with a bulk
scalar $\phi$ when singular branes are absent:
\begin{equation}
\oint dy \; \phi^{\prime}\cdot\phi^{\prime}=0.
\label{GKL_sr}
\end{equation}
The above result implies that non-trivial scalar profiles are inconsistent with periodicity, the only
allowed configuration is $\phi=$const..
The sum rule (\ref{GKL_sr}) was obtained assuming minimal scalar-gravity coupling. In the following we are
going to generalize the result for the case of non-minimal coupling described by the action (\ref{action_m}).

It is easy (subtracting Eqs. \eqref{eom01_m} and \eqref{eom02_m})
to derive an equation of motion that contains only the warp function $A(y)$ and the input profile
$\phi(y)$:
\beq
3fA^{\prime\prime} = f^{\prime} A^{\prime} - f^{\prime\prime} - \frac{1}{2}(\phi^{\prime})^{2}.
\label{eom01a_m}
\eeq
It is useful to rewrite the above equation by the change of variables $X(y)=A^{\prime}(y)$:
\beq
X^{\prime}(y)=F(y)X(y)+G(y),
\label{eom03a_m}
\eeq
where,
\begin{align}
F(y)&\equiv \frac{f^{\prime}(y)}{3 f(y)},
\label{eom04a_m}\\
G(y)&\equiv -\frac{f^{\prime\prime}(y)+\frac{1}{2} \phi^{\prime}(y)^2}{3 f(y)}.
\label{eom05a_m}
\end{align}
We assume that the profile is periodic with a period $L$:
\[
\phi(y+L)=\phi(y),
\]
then $f(\phi)$ and consequently, $F(y)$ and $G(y)$ are also periodic with the same period $L$.
Since
\beq
\oint dy\;F(y)=0
\label{con1}
\eeq
it is straightforward to notice that the solution of the homogeneous part of Eq.~\eqref{eom03a_m}
\beq
X(y)=X_0 e^{\int_{y_0}^y F(s)ds}
\label{sol_X}
\eeq
is periodic as well.

The inhomogeneous equation \eqref{eom03a_m} could be rewritten in the following form
\beq
\left[Z(y)X(y)\right]^{\prime}=Z(y)G(y)
\label{ode3},
\eeq
where
$
Z(y)\equiv Z(0)e^{-\int_{0}^{t} F(s)ds}
$
is a solution of the following homogeneous equation,
\begin{align}
Z^{\prime}(y)&=-F(y)Z(y).
\label{ode2}
\end{align}
Integrating \eqref{ode3} over the period we obtain the following condition:
\beq
\oint dy \;G(y)Z(y)=0,
\label{GKL_rg_gen}
\eeq
which constitutes the proper generalization of the Gibbons-Kallosh-Linde sum rule (\ref{GKL_sr}).
For $F(y)$ defined in (\ref{eom04a_m}) one obtains $Z(y)$ explicitly
\beq
Z(y)=Z(0)\left[\frac{f(0)}{f(y)}\right]^{1/3}
\label{Z}
\eeq
Then, after integrating by parts, the sum rule (\ref{GKL_rg_gen}) reads
\beq
\oint dy\; \left[\frac43 \frac{1}{f^{1/3}} \left(\frac{f^\prime}{f}\right)^2 + \frac12 \left( \frac{\phi^{\prime}}{f^{2/3}}\right)^2\right] = 0
\label{sum_rule}
\eeq
The above sum rule again implies that even in the presence
of non-minimal couplings, $f(\phi) R$, only the trivial profile, $\phi=$const., is consistent with
periodicity.

Note that \eqref{sum_rule} holds also for multicomponent scalar fields, therefore even
in that case non-trivial profiles in the absence of singular branes are excluded by periodicity.
It is also worth mentioning that the result (\ref{sum_rule}) could be obtained
by writing the action \eqref{action_m} in the Einstein frame where the scalar field is
minimally coupled to gravity. In the Einstein frame the standard GKL sum-rule \eqref{GKL_sr}
holds, and when the sum rule is rewritten back in the Jordan frame defined by \eqref{action_m},
the result \eqref{sum_rule} is reproduced.

\subsection{Negative tension brane in modified gravity}
\label{Negative tension brane in modified gravity}

In the RS1 set-up one of the two 3-branes must have a negative tension. Therefore in this subsection
we turn to the question weather a negative tension brane can be constructed out of a real
scalar field in the modified gravity scenario when the scalar field is non-minimally coupled
to gravity as in (\ref{action_m}). By the virtue of the result of the previous subsection we discard
the possibility of periodic extra dimensions. Let us assume, without loosing any generality,
that the scalar filed $\phi(y)$ has a kink-like profile,
\beq
\phi(y)=\frac{\kappa}{\sqrt{\beta}} \tanh (\beta y),
\label{nb1}
\eeq
where $\beta$ is a brane-thickness controlling parameter. For $\beta \to \infty$ (the brane limit), as
it will be discussed in details in
Sec.~\ref{The brane limit}, the profile $\phi(y)$ generates singular energy density localized at
$y=0$ that could mimic a D3-brane.

The action for the kink configuration (\ref{nb1}) can be written as,
\beq
S_{\phi}=-\int d^{5}x \sqrt{-g}\left[\frac{1}{2}(\phi^{\prime})^2+V(\phi)\right],
\label{nb2}
\eeq
while the action for a brane localized at $y=0$ with a negative tension ($\lambda>0$) reads
\beq
S_{D3}=\int d^{5}x \sqrt{-g}\lambda \delta(y).
\label{nb3}
\eeq
Using the equations of motion \eqref{eom01_m} and \eqref{eom02_m} we can rewrite the action \eqref{nb2} as follows
\beq
-\int dy \left[\frac{1}{2}(\phi^{\prime})^{2}+V\right] =
\int dy \left[-(\phi^{\prime})^{2} + 12(A^{\prime})^{2}f + 8 A^{\prime}f^{\prime} \right].
\label{nb4}
\eeq
As it will be clear from the next section, the only interesting set-up is such that the warp function
$A(y)$ reaches its maximum at the brane location (so $A^\prime(0)=0$), therefore among the above terms
only the very first one contributes to the brane tension. However, as it is seen from \eqref{nb3} there
is no possibility to reproduce the sign required by the negative tension.
Therefore we conclude that a single kink-like profile can generate only a positive tension brane
even in the case of modified gravity.

\subsection{Conclusions on thick brane generalization of the RS1 model}
\label{Conclusions on thick brane generalization of the RS1 model}

As we have shown in the proceeding subsections there is a conflict between the RS1
scenario and the idea of branes generated by bulk scalar profiles:
\bit
\item As shown in Sec.\ref{Thick branes with periodic extra dimensions}, even in
the presence of the non-minimal scalar-gravity coupling $f(\phi)R$, periodicity
in the extra coordinate can not be reconciled with a non-trivial profile.
\item One of the branes in the RS1 scenario must have negative tension, however as
we have shown in Sec.\ref{Negative tension brane in modified gravity} even if scalars
interact non-minimally with gravity
there is no way to generate a brane with negative tension.
\eit
The above observations prompt to give up compactness and therefore to discuss
a possibility of mimicking the RS2 model with non-compact extra dimension. Since we would like to
allow for the solution of the hierarchy problem by the virtue of warping the metric along extra dimensions,
we will introduce a scalar field, the profile of which could mimic a scenario with two branes
of positive tension with AdS metric in between them. This is what we are going to discuss in the
following sections limiting ourselves to the case of minimal scalar-gravity coupling, however the
analysis could be easily extended to non-minimal scenarios as well.

\section{Two thick branes and the background solutions}
\label{Two thick branes and the background solutions}

From now onwards we will adopt the following action for a 5D scalar field minimally coupled
to the Einstein-Hilbert gravity
\begin{equation}
{\cal S}=\int dx^5 \sqrt{-g}\left\{2M^{3}R-\frac{1}{2}g^{MN}\nabla_{M}\phi\nabla_{N}\phi-V(\phi)\right\},
\label{action}
\end{equation}
where $g_{MN}$ is the warped 5D metric,
\begin{equation}
g_{MN}=\left( \begin{array}{cc}e^{2A(y)}\eta_{\mu\nu} & 0\\ 0& 1\end{array}\right).
\label{G_MN}
\end{equation}
The Einstein's equations and the equation of motion for $\phi$, resulting from the action
\eqref{action} are
\begin{eqnarray}
R_{MN}-\frac{1}{2}g_{MN}R&=&\frac{1}{4M^{3}}T_{MN},\label{eineq1}\\
\nabla^{2}\phi-\frac{dV}{d\phi}&=&0, \label{eineq2}
\end{eqnarray}
where $\nabla^2$ is 5D covariant d'Alambertion operator (see Appendix-\ref{Conv} for conventions and details)
and the energy-momentum tensor $T_{MN}$ for the scalar field $\phi(y)$ is,
\begin{equation}
T_{MN}=\nabla_{M}\phi\nabla_{N}\phi-g_{MN}\left[\frac{1}{2}(\nabla\phi)^{2}+V(\phi)\right].\label{emt}
\end{equation}
From the Einstein equations \eqref{eineq1} and \eqref{eineq2}, one can get the following equations of motion
for the metric ansatz \eqref{G_MN}
\begin{align}
24M^{3}(A^{\prime})^{2}&=\frac{1}{2}(\phi^{\prime})^{2}-V(\phi),\label{eom01}\\
12M^{3}A^{\prime\prime}+24M^{3}(A^{\prime})^{2}&=-\frac{1}{2}(\phi^{\prime})^{2}-V(\phi),\label{eom02}\\
\phi^{\prime\prime}+4A^{\prime}\phi^{\prime}&-\frac{dV}{d\phi}=0. \label{eom03}
\end{align}
We assume that the scalar potential $V(\phi)$ could be expressed
in terms of the superpotential $W(\phi)$ as \cite{DeWolfe:1999cp},
\begin{equation}
V(\phi)=\frac{1}{2}\left( \frac{\partial W(\phi)}{\partial \phi}\right)^{2}-\frac{1}{6M^{3}}W(\phi)^{2},
\label{potential}
\end{equation}
where $W(\phi)$ satisfies the following relations,
\begin{equation}
 \phi^{\prime}=\frac{\partial W(\phi)}{\partial \phi} \hspace{1cm}\text{and}\hspace{1cm} A^{\prime}=-\frac{1}{12M^{3}} W(\phi).
 \label{super_potential}
\end{equation}
Although the use of this method is motivated by supersymmetry, no supersymmetry is involved in
our set-up. The method is elegant and
very efficient, in particular it implies that the system of second order differential equations
\eqref{eom01}-\eqref{eom03} is now
reduced to first order ordinary differential equations which are much easier to deal with. We
also point out that the superpotential
method is fully equivalent to the standard approach (i.e. solving the equations of motion) as
long as the solutions for scalar
field have monotonic profile, as it will be the case in our model with kink-like profiles.
It is worth to mention that the standard (and straightforward) application of the superpotential
method is limited to the single scalar-field case since with multi-scalar fields it
becomes difficult to handle analytically.

We are interested in the case where the scalar field $\phi(y)$ is given by a kink-like profiles~\footnote{The
profile of the scalar field could be different than the standard kink but the essential concept
holds for any profile which is monotonic and satisfies equations of motion.}, i.e.,
\begin{equation}
 \phi(y)=\sum_{\alpha} \frac{\kappa_{\alpha}}{\sqrt{\beta_{\alpha}}}\tanh(\beta_{\alpha}(y-y_{\alpha})),
\label{scalar}
\end{equation}
where $\beta_{\alpha}$ are the thickness regulators and $\kappa_{\alpha}$ parameterize tensions
of the branes in the
so called \emph{brane limit} $\beta_{\alpha}\to \infty$ (from here on we will consider
$\beta_{\alpha}=\beta$, i.e.
all the branes have equal thickness, although that could be relaxed).
As it will be shown in the next section the profile (\ref{scalar}) in the brane limit corresponds
to 3-branes with brane-tensions given by
\begin{equation}
\lambda_{\alpha}=\frac{4}{3}\kappa_{\alpha}^{2}. \label{brane_tension}
\end{equation}
It is important to note that this set-up implies that only positive brane tensions could be
mimicked by scalar
filed configurations, as was also pointed out by DeWolfe et al. \cite{DeWolfe:1999cp}.
Therefore the scalar field
can not reproduce the RS1 scenario where the IR brane has a negative tension.

In our set-up we consider two kinks corresponding to two thick branes at locations $y=y_1$
and $y=y_2$.
They are supposed to mimic two positive-tension branes in the brane limit, so the scalar
profile $\phi(y)$ could be chosen as follows,
\begin{equation}
 \phi(y)= \frac{\kappa_{1}}{\sqrt{\beta}}\tanh(\beta(y-y_{1}))+\frac{\kappa_{2}}{\sqrt{\beta}}\tanh(\beta(y-y_{2})).
\label{scalar_kinks}
\end{equation}
We can find the superpotential $W(\phi)$ in such a way that it allows a solution of the scalar
field $\phi(y)$ as in
Eq.~\eqref{scalar_kinks}.
This can be obtained from Eq.~\eqref{super_potential} as,
\begin{align}
\phi^{\prime}(y)&=\frac{\partial W(\phi)}{\partial\phi}=\frac{\partial W(\phi(y))}{\partial y}\frac{\partial y}{\partial\phi(y)}=
\frac{W^{\prime}(y)}{\phi^\prime(y)},\label{super_potential_1}\\
W(y)&=\int (\phi^{\prime}(y))^{2}dy+W_{0},
\label{super_potential_2}
\end{align}
where $W_0$ is some constant of integration. Deriving the above relation it is assumed that
$\phi(y)$ is an invertible
function of $y$, so that the superpotential could be written as a function of $y$ as
\begin{align}
W(y)=&\kappa_1^2\left\{\tanh[\beta(y-y_{1})]-\frac{1}{3}\tanh^{3}[\beta(y-y_{1})]\right\}\notag\\
&+\kappa_2^2\left\{\tanh[\beta (y-y_{2})]-\frac{1}{3}\tanh^{3}[\beta (y-y_{2})]\right\}+W_{0},
\label{super_potential_brane_1}
\end{align}
where in deriving Eq.~\eqref{super_potential_brane_1} we assume that the cross term is
negligible as far as $\beta$
is large and/or the separation $``y_2-y_1"$ between the two thick branes is large such that,
\[
\int dy \frac{2\beta\kappa_1\kappa_{2}}{\cosh^2(\beta ( y-y_1))\cosh^2(\beta ( y-y_2))}\approx 0
\]
After obtaining the superpotential $W(y)$ we find from Eq.~\eqref{super_potential},
\begin{align}
A^{\prime}(y)=-\frac{1}{12M^{3}}&\left\{\kappa_1^2\left(\tanh(\beta(y-y_{1}))-\frac{1}{3}\tanh^{3}(\beta(y-y_{1}))\right)\right.\notag\\
&\left.+\kappa_2^2\left(\tanh(\beta (y-y_{2}))-\frac{1}{3}\tanh^{3}(\beta (y-y_{2}))\right)+W_{0}\right\}.
\label{A_prime}
\end{align}
The integration constant $W_{0}$ can be fixed by the requirement that $A(y)$ has a maximum at $y=y_0$.
The location of maximum with respect to $y_{1,2}$ will correspond to different 5D geometric configurations
that we will discuss in
Sec.\ref{The brane limit}. Therefore, we choose the integration constant $W_{0}$ as,
\begin{align}
W_{0}&=-\left\{\kappa_1^2\left(\tanh(\beta (y_{0}-y_{1}))-\frac{1}{3}\tanh^{3}(\beta (y_{0}-y_{1}))\right)\right.\notag\\
&\left.+\kappa_2^2\left(\tanh(\beta (y_{0}-y_{2}))-\frac{1}{3}\tanh^{3}(\beta (y_{0}-y_{2}))\right)\right\},
\label{W_0}
\end{align}
such that $A'(y_0)=0$.

Now it is straightforward to find the warp factor $A(y)$ by integrating Eq.~\eqref{A_prime} w.r.t. $y$.
The result reads,
\begin{align}
A(y)=&\frac{1}{72M^{3}\beta}\left\{\kappa_1^2\left(\frac{1}{\cosh^{2}(\beta(y-y_{1}))}-\ln\cosh^{4}(\beta(y-y_{1}))\right)\right.\notag\\
&\left.+\kappa_2^2\left(\frac{1}{\cosh^{2}(\beta (y-y_2))}-\ln\cosh^{4}(\beta (y-y_2))\right)\right\}+\frac{1}{12M^{3}}W_{0}y+A_{0},
\label{warp-factor}
\end{align}
where $A_{0}$ is a constant of integration which can be fixed by the requirement such that $A(y_0)=0$.
Note that far away from the thick branes the warp function takes the RS form
\cite{Randall:1999ee, Randall:1999vf}
\begin{align}
A(y\to \infty)&\sim -\kappa\vert y\vert, \hspace{2cm} |y|\gg|y_1-y_2|,
\label{A_prime_RS}
\end{align}
where $\kappa=\frac{1}{24M^{3}}\left(\frac{4}{3}\kappa_1^2+\frac{4}{3}\kappa_2^2-W_{0}\right)$.
It is easy to see that we get the same behavior, i.e., the RS form of $A(y)$, for all
values of $y$ in the brane limit when $\beta\to\infty$, which will be discussed in detail in the next section.

It should be mentioned that this set-up reduces to the usual single thick brane \cite{Kehagias:2000au}
if we assume that one
of the branes is far away from the other brane, say $y_1=0$ and $y_2\to\infty$ such that the second brane
can be removed from
the set-up and in that case $\kappa=\frac{1}{24M^{3}}\left(\frac{4}{3}\kappa_1^2\right)$.
Unfortunately in the one-brane case the hierarchy problem remains unsolved, that is why we focus on
the two-brane scenario
as the hierarchy problem is one of the main motivations of this work. In our case the second thick
brane is located in the AdS space at suitable location $y_2$ such that the hierarchy problem could
be addressed as in a similar way as it was done in the original RS1 set-up \cite{Randall:1999ee} or
in the LR set-up \cite{Lykken:1999nb}.

\begin{figure*}[ht]\centering
\includegraphics[scale=0.7]{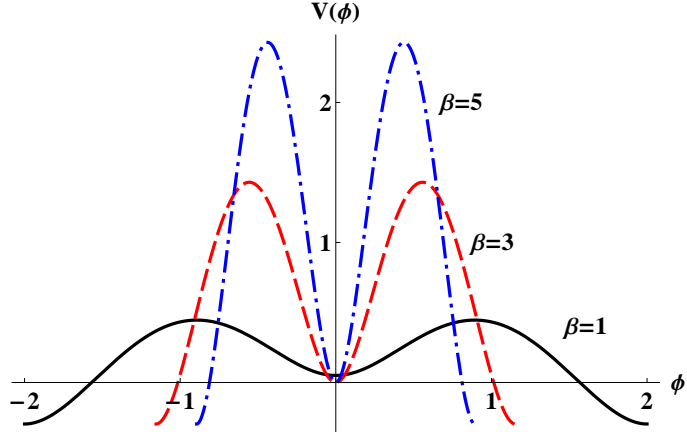}
\caption{The potential $V(\phi)$ plotted as a function of the scalar field $\phi$ for different values of the
thickness parameter $\beta$ with the same brane tensions $\kappa_1=\kappa_2=1$. Hereafter we
assume $M=1$, therefore the field strength is expressed in unites of $M^{3/2}$.}
\label{Vofphi}
\end{figure*}

Before closing this section it is instructive to discuss the shape of the scalar potential, that
is determined by
our requirement of having (\ref{scalar_kinks}) together with the ansatz (\ref{G_MN}) as solutions
of the equations of motion.
Having the superpotential $W$ determined, one can, using \eqref{potential}, find the scalar
potential as a function of $y$. However, since $\phi(y)$ is an invertible function, therefore it is also
possible to plot the potential $V(\phi)$ (\ref{potential}), as a function of $\phi$.
However, in order to develop some intuition,
let us first consider the presence of just one-kink profile (\ref{nb1}) with parameters $\kappa$ and $\beta$.
Then the potential  $V(\phi)$ could be determined analytically since one can easily solve equations of motion
for $A(y)$, from the invertible profile one can find $y=y(\phi)$ and then adopt it, for instance, in (\ref{eom01}). The results reads:
\beq
V(\phi)=\frac{\beta^3}{2\kappa^2}\left(\phi^2-\frac{\kappa^2}{\beta}\right)^2 -\frac{1}{54M^3}\frac{\beta^3}{\kappa^2}\phi^2\left(\phi^2-3\frac{\kappa^2}{\beta}\right)^2.
\label{pot1kink}
\eeq
Note that the above form of the potential applies, strictly speaking, only for
$-\kappa/\sqrt{\beta}<\phi<\kappa/\sqrt{\beta}$ since the profile that is required
to fulfill equations of motion varies in that range. For small field strengths, $\phi \lsim M^{3/2}$,
gravitational effects (the second term) are small, so that the dominant contribution is just the
Mexican-hat potential (the first term).
In the real case the scalar profile is a sum of two kinks (\ref{scalar_kinks}),
therefore its range of variation is roughly a ``sum" of ranges for two separate kinks. In the absence of gravity each kink is a solution of equation of motion for a Mexican-hat type potential, again if a field strength is small comparing to the 5D
Planck mass$^{3/2}$ the dominant contribution to the potential is the Mexican-hat component. Since kinks profiles
vary in between field strengths corresponding to the two minima therefore the shapes which
we observe  in Fig. \ref{Vofphi} are roughly ``sums" of inner parts (only the inner part could be determined) of two
Mexican-hat like potentials; one centered around $-\kappa_1/(2 \sqrt{\beta})$ and the other one around $+\kappa_2/(2 \sqrt{\beta})$. In reality (with gravity) the picture is slightly distorted by the gravity effects that become relevant around the external ends of region of variation where $\phi/M^{3/2}\sim 1$.
\newline In the case of double kink it is not possible to find the potential analytically, so $V(\phi)$
determined numerically is shown for several choices of the parameters $\beta$ in Fig.~\ref{Vofphi}.
Since the strength of the
profile field varies between $-(\kappa_1+\kappa_2)/\sqrt{\beta}$ and  $+(\kappa_1+\kappa_2)/\sqrt{\beta}$
therefore the potential $V(\phi)$ can also be determined in that region only, which is manifest in Fig.~\ref{Vofphi}.
Note that in order to trust classical field theory results the scalar field strength $\phi$ must be limited by the
5D Planck mass $M^{3/2}$, therefore we conclude that our results are consistent if $\beta \gsim k_i^2/M^3$. Since in
Fig.~\ref{Vofphi} we assumed $\kappa_1=\kappa_2=1$ and $M=1$ therefore we are limited by $\beta \gsim 1$, so those cases were plotted and then the range of variation of $\phi$ is appropriate.

\section{The brane limit}
\label{The brane limit}

In this section we will consider different possible scenarios that could be realized with two thick branes and then
we will discuss limiting (the brane limit) solutions corresponding to thin (singular)
branes. To show how scalar field $\phi$ is making a brane, we start by looking at its action calculated for the
profile (\ref{scalar_kinks}),
\begin{align}
{\cal S}_{\phi}&=\int dx^5 \sqrt{-g}\left\{-\frac{1}{2}g^{MN}\nabla_{M}\phi\nabla_{N}\phi-V(\phi)\right\},\notag\\
&=\int dx^5 \sqrt{-g}\left\{-(\phi^\prime)^{2}+\frac{1}{6M^{3}}W^{2}(\phi) \right\},\notag\\
&=\int dx^5 \sqrt{-g}\left\{-\frac{\beta\kappa_1^2}{\cosh^4(\beta ( y-y_1))}-\frac{\beta\kappa_2^2}{\cosh^4(\beta ( y-y_2))} \right.\notag\\
&\hspace{2cm}\left.+\frac{1}{6M^{3}}\left[\kappa_1^2\left(\tanh(\beta ( y-y_1))-\frac{1}{2}\tanh^{3}(\beta ( y-y_1))\right)\right.\right.\notag\\
&\hspace{2cm}\left.\left.+\kappa_2^2\left(\tanh(\beta (y-y_{2}))-\frac{1}{3}\tanh^{3}(\beta (y-y_{2}))\right) +W_{0} \right]^{2}\right\}. \label{action-xi}
\end{align}
In the brane limit we have\footnote{Note that we have corrected the misprint that appeared in \cite{Kehagias:2000au} in their footnote 3.},
$$
\lim_{\beta\to\infty}\left\{\frac{\beta}{\cosh^4(\beta ( y-y_i))}\right\}=\frac{4}{3}\delta(y-y_i)\lsp {\rm for} \lsp i=1,2
$$
such that the scalar action \eqref{action-xi} can be written as,
\begin{align}
{\cal S}_{\phi}
&=\int dx^5 \sqrt{-g}\left\{-\frac{4}{3}\kappa_1^2\delta(y-y_1)-\frac{4}{3}\kappa_2^2\delta( y-y_2)-\Lambda_B(y)\right\},\label{action-xi1}
\end{align}
where $\Lambda_B(y)$ is a function that generates cosmological constants in various regions of the bulk:
\begin{align}
\Lambda_B(y)&=\lim_{\beta\to\infty}\bigg[-\frac{1}{6M^{3}} \left\{\kappa_1^2\left(\tanh(\beta ( y-y_1))-\frac{1}{3}\tanh^{3}(\beta ( y-y_1))\right)\right.\notag\\
&\hspace{2cm}\left.+\kappa_2^2\left(\tanh(\beta (y-y_{2}))-\frac{1}{3}\tanh^{3}(\beta (y-y_{2}))\right) +W_{0} \right\}^{2}\bigg],\notag\\
&=-\frac{1}{6M^{3}} \left\{\frac{2}{3}\kappa_1^2 sgn( y-y_1))+\frac{2}{3}\kappa_2^2 sgn( y-y_2))+W_{0}\right\}^{2}.\label{lambda-B}
\end{align}
Therefore, depending on the choice of the extremum location $y_{0}$, different values of cosmological constant to the left, in between and
to the right of the two branes could be generated.

In what follows we will find analytic solutions for the two positive tension branes in the brane limit $(\beta\to\infty)$
and numerical results for the corresponding thick brane scenarios. In the brane limit, we can write the total action as,
\begin{align}
{\cal S}_{BL}
&=\int dx^5 \sqrt{-g}\left\{2M^{3}R-\lambda_{1}\delta(y-y_1)-\lambda_{2}\delta( y-y_2) -\Lambda_{B}(y)\right\},\label{action-bl}
\end{align}
where $\lambda_{1,2}=\frac{4}{3}\kappa_{1,2}^2$ are the respective brane tensions at each brane located at $y=y_1$ and $y=y_2$ and $\Lambda_{B}(y)$
is the bulk cosmological constant, defined in Eq.~\eqref{lambda-B}. In the brane limit we can obtain the equations of motion from action \eqref{action-bl} as,
\begin{align}
24M^{3}\left(A^{\prime}\right)^{2}&=-\Lambda_{B},\label{aeom01}\\
12M^{3}A^{\prime\prime}+24M^{3}\left(A^{\prime}\right)^{2}&=-\Lambda_{B}
-\lambda_{1}\delta(y-y_{1})-\lambda_{2}\delta(y-y_{2}),\label{aeom02}
\end{align}
In the brane limit the smooth solution of $A^{\prime}(y)$ \eqref{A_prime} will take the following form,
\begin{align}
A^{\prime}(y)=-\frac{1}{12M^{3}}&\left\{\frac{2}{3}\kappa_1^2 sgn(y-y_{1})+\frac{2}{3}\kappa_2^2 sgn(y-y_{2})+W_{0}\right\}.\label{A_prime1}
\end{align}
From Eqs. \eqref{aeom01} and \eqref{aeom02}, one gets (which is also manifested from Eq.~\eqref{A_prime1}),
\begin{align}
12M^{3}A^{\prime\prime}&=
-\lambda_{1}\delta(y-y_{1})-\lambda_{2}\delta(y-y_{2}),\label{aeom03}
\end{align}
which implies that $A^{\prime\prime}<0$ as $\lambda_{1,2}>0$, thus alowing for maxima of $A(y)$.
Let us denote the location of the maxima by $y_0$, that can be chosen anywhere along the extra-dimension, such that,
\begin{align}
A^{\prime}(y_{0})&=0.\label{max1}
\end{align}
In the brane limit $W_{0}$ \eqref{W_0} takes the form,
\begin{align}
W_{0}&=-\left\{\frac{2}{3}\kappa_1^2 sgn(y_{0}-y_{1})+\frac{2}{3}\kappa_2^2 sgn(y_{0}-y_{2})\right\}.\label{W_0_1}
\end{align}
Therefore we obtain (in the brane limit) for the bulk cosmological constant, the following result
\beq
\Lambda_B(y)=-\frac{1}{6M^{3}} \left\{\frac{2}{3}\kappa_1^2 sgn( y-y_1)+\frac{2}{3}\kappa_2^2 sgn(y-y_2)-\frac{2}{3}\kappa_1^2 sgn(y_{0}-y_{1})-\frac{2}{3}\kappa_2^2 sgn(y_{0}-y_{2})\right\}^{2}. \label{lamB}
\eeq
It is also important to note that the equation of motion \eqref{aeom01} implies that the bulk cosmological constant is
negative
leading to anti-de Sitter vacua or in the case where it is zero that the corresponding geometry will be Minkowski in that region of space.

In the following we will consider different cases depending on the location of the extremum point $y_{0}$ along the extra-dimension.
\newline\textbf{Case-I:} We consider the case when the extremum location is on one of the branes, say at $y_{0}=y_{1}$, we get the
analytic results for $A^{\prime}(y)$ as,
\begin{align}
A^{\prime}(y)=\left\{\begin{array}{l c}\frac{1}{24M^{3}} \lambda_{1}\hspace{2cm}& y < y_{1}\\
-\frac{1}{24M^{3}} \lambda_{1} &y_{1}<y < y_{2}\\
-\frac{1}{24M^{3}} \left( \lambda_{1} +2 \lambda_{2}\right) &y > y_{2}\end{array}\right.
\label{Ap4}
\end{align}
The corresponding bulk cosmological constant $\Lambda_{B}$, in different regions along the extra-dimension, reads as,
\begin{align}
\Lambda_{B}
=\left\{\begin{array}{l c} -\frac{1}{24M^{3}}\lambda_{1}^{2} \hspace{2cm}&y < y_{1}\\
 -\frac{1}{24M^{3}} \lambda_{1}^{2} &y_{1}< y < y_{2}\\
 -\frac{1}{24M^{3}} \left(\lambda_{1}+2\lambda_{2}\right)^{2}&y > y_{2}\end{array}\right. ,
 \label{LamB3}
\end{align}
and the values of bulk cosmological constant at the brane locations are $\Lambda_B(y_1)=0$ and $\Lambda_B(y_2)=-\frac{1}{12M^{3}} \left(2\lambda_{2}^{2}+\lambda_{1}\lambda_{2}\right)$.
Note that (\ref{LamB3}) implies in the brane limit correlations between the brane tensions $\lambda_i$ and the bulk
cosmological constant $\Lambda_B$. It is worth to write down relations between $\Lambda_B$,  the slope of the warp function
$A^\prime$ and the brane tensions $\lambda_i$ in the asymptotic regions:
\begin{align}
\frac{\Lambda_B}{A^\prime} = \left\{
\begin{array}{l c}
-\lambda_1 & \lsp y \to -\infty\\
\lambda_1+2\lambda_2 & \lsp y \to +\infty
\end{array}
\right.
\label{asym}
\end{align}
In the RS2 (with one brane of positive tension $\lambda$) the corresponding relation is $\Lambda_B/k=\pm \lambda$, where
$k$ corresponds to $\pm A^\prime$.

\begin{figure*}[ht]
\begin{tabular}{cc}
($a$)&($b$)\\
\includegraphics[scale=0.55]{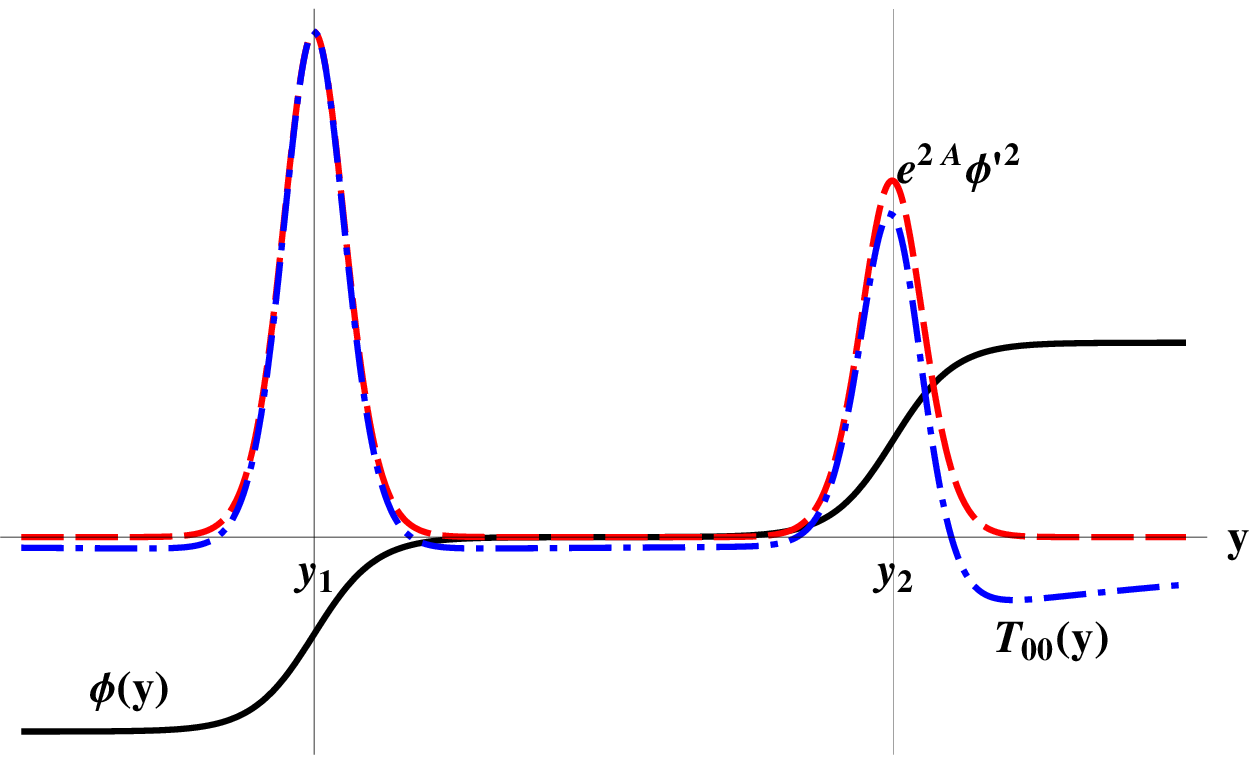} \
& \includegraphics[scale=0.55]{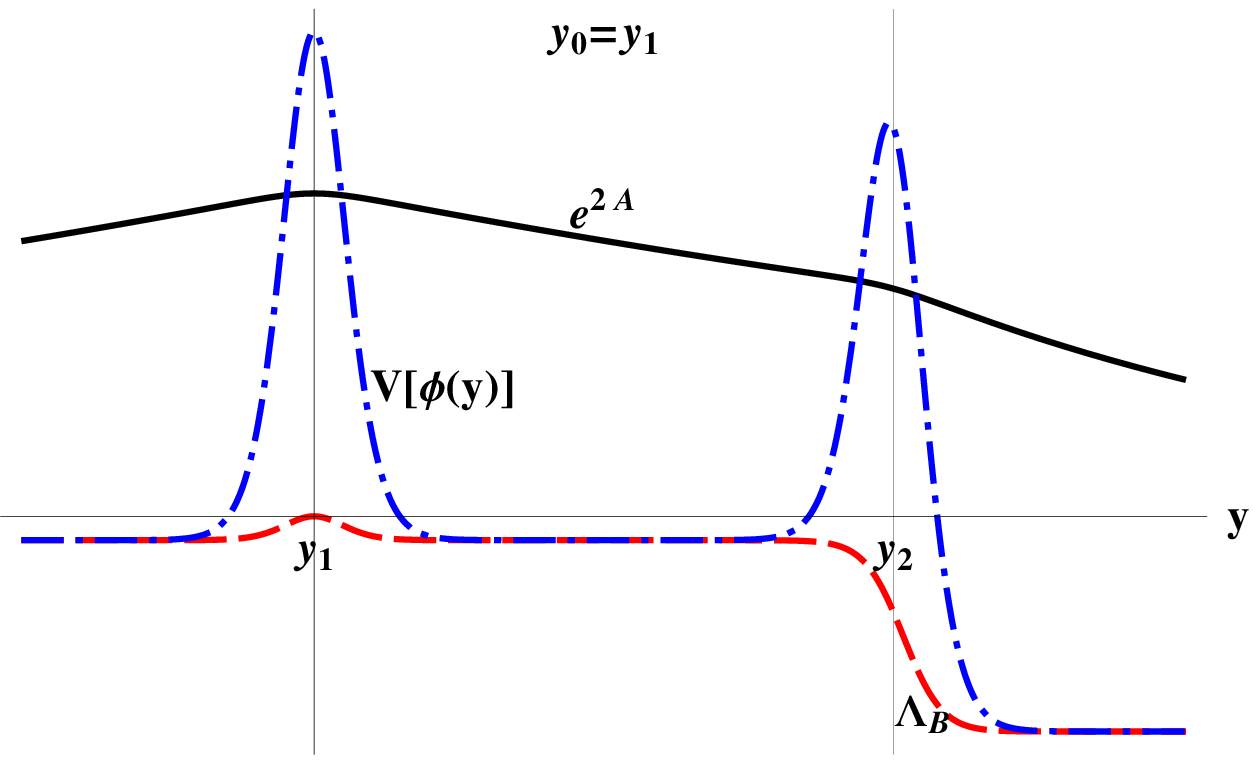}
\end{tabular}
\caption{Graph $(a)$ shows the shape of the scalar field profile, the corresponding energy density $T_{00}$ and also
the localized contribution to $T_{00}$ (i.e. $e^{2A}\phi^{\prime\,2}$) for the Case-I with $\beta=3$ and $\kappa_1=\kappa_2=1$,
whereas, graph $(b)$ illustrates the shape of the warped factor $e^{2A}$, the bulk cosmological constant $\Lambda_B$ and
the potential $V[\phi(y)]$ for the case-I when the maxima of the warp factor is located on the thick brane at $y_1$.
}
\label{fig1}
\end{figure*}

Numerical solutions for the thick branes in the Case I are shown in Fig.~\ref{fig1}. This configuration is such that the
warping function $A(y)$ is positive (negative) to the left (right) of the branes, we will see that this scenario
will have the normalizable zero-modes of the metric tensor perturbations (around the background solution)
which correspond to the 4D graviton.
In the brane limit this set-up is similar to the two
positive D3-brane model considered by Lykken and Randall in \cite{Lykken:1999nb}.

In order to illustrate how are the branes generated, in Fig.~\ref{fig1}$(a)$ we show the energy density
$T_{00}=e^{2A}(\phi^{\prime\,2}/2+V(\phi))$ corresponding to the profile. Using equations of motion
$T_{00}$ could be rewritten as $e^{2A}(\phi^{\prime\,2}-24 M^3 A^{\prime\,2})$. This form separates two contributions
to $T_{00}$: local (one that "creates" the branes) $e^{2A}\phi^{\prime\,2}$ and non-local
$\propto e^{2A} A^{\prime\,2}$ (one that generates the bulk cosmological constants).

As it is seen from Fig.~\ref{fig1}$(b)$ to the left and to the right of the branes the warp factor is quickly vanishing,
that has been already observed in (\ref{A_prime_RS}). If the branes are sufficiently thin
(or well separated) then in between them the warping is also nearly exponential so that
the hierarchy problem could be addressed. We will call the brane located at $y_1$ and $y_2$ as UV and IR branes,
respectively. To illustrate consequences of the warped background geometry lets assume that the
Higgs field is bounded at the IR brane and its action can be written as,
\begin{align}
{\cal S}_H&= -\int d^4x\sqrt{-\hat g}\left\{\hat g_{IR}^{\mu\nu} \partial_\mu H^\dagger \partial_\nu H-m^2|H|^2+\lambda |H|^4 \right\}, \label{Higgs_1}
\end{align}
where $\hat g_{IR}^{\mu\nu}$ is the 4D metric induced on the IR brane, $\hat g_{IR}^{\mu\nu}=e^{-2A(y_2)}\eta^{\mu\nu}$,
with $A(y_2)$ being the value of warped factor at the IR brane and $m$
is the 5D Higgs mass parameter (of the order of 5D Planck mass). Now the effective 4D action for the Higgs
field can be written as,
\begin{align}
{\cal S}_H&= -\int d^4x\left\{e^{2A(y_2)}\eta^{\mu\nu} \partial_\mu H^\dagger \partial_\nu H-m^2e^{4A(y_2)}|H|^2+\lambda e^{4A(y_2)}|H|^4 \right\}, \label{Higgs_2}
\end{align}
where we used the fact that, $\sqrt{-\hat g}=e^{4A(y_2)}$. In order to obtain canonically normalized Higgs field,
we rescale, $H\to e^{-A}H$, such that,
\begin{align}
{\cal S}_H&= -\int d^4x\left\{\eta^{\mu\nu} \partial_\mu H^\dagger \partial_\nu H-m^2e^{2A(y_2)}|H|^2+\lambda |H|^4 \right\}, \non \\
&=-\int d^4x\left\{\eta^{\mu\nu} \partial_\mu H^\dagger \partial_\nu H-\mu^2|H|^2+\lambda |H|^4 \right\},
\label{Higgs_4}
\end{align}
where $\mu=me^{A(y_2)}$ is the effective Higgs mass parameter as viewed on the IR brane. If we assume that the
fundamental mass scale of the
5D theory is the Planck mass then we can require the value of warped factor at IR brane such that we get the
effective 4D Higgs mass parameter $\mu \sim \tev$. From Eq.~\eqref{Ap4} we have
\begin{align}
A(y_2)=-\frac{1}{24M^{3}} \left( \lambda_{1} + \lambda_{2}\right) y_{2},     \label{A_brane_limit}
\end{align}
therefore if $\frac{1}{24M^{3}} \left( \lambda_{1} + \lambda_{2}\right)y_2 \sim 30$ then the hierarchy
problem could be solved. Furthermore,
as it will be shown in Sec. \ref{Stability of tensor perturbations}, in this scenario (i.e. $y_0=y_1$)
there exits a normalizable zero-mode which corresponds to the 4D graviton.

\begin{figure*}[ht]
\begin{tabular}{cc}
($a$)&($b$)\\
\includegraphics[scale=0.55]{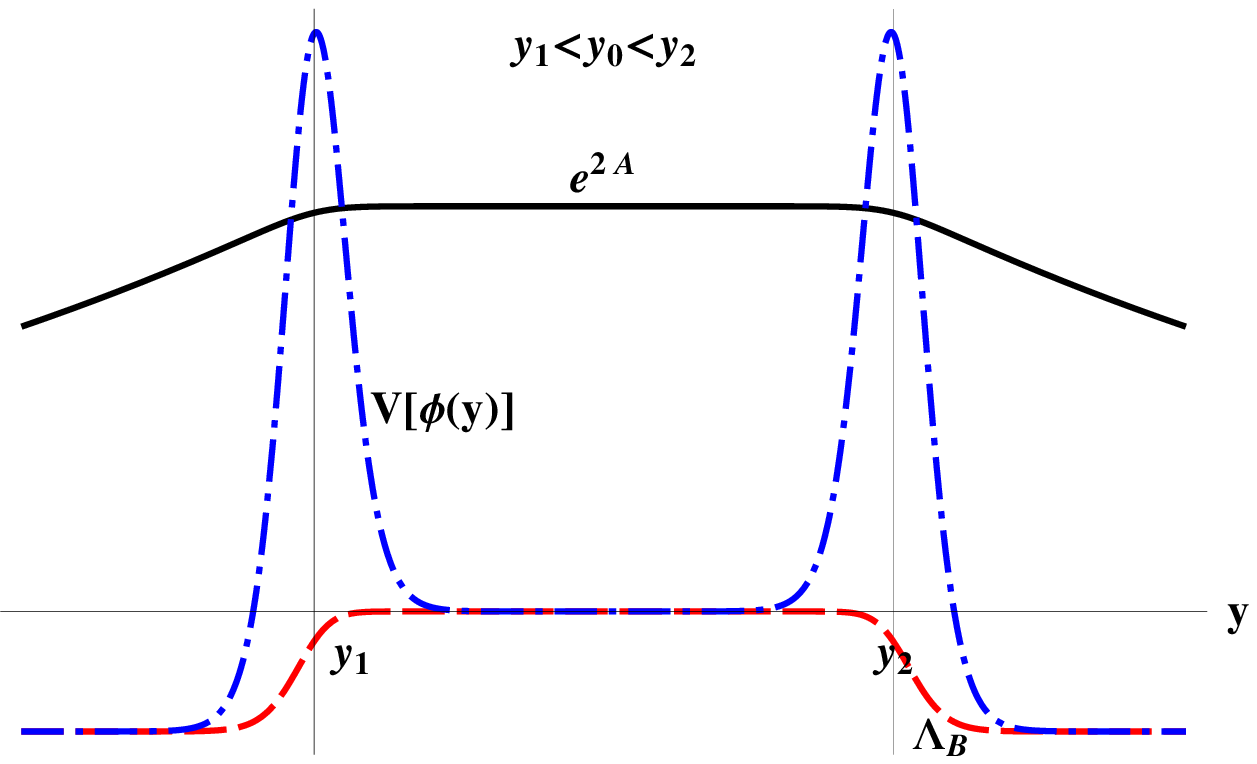} \
& \includegraphics[scale=0.55]{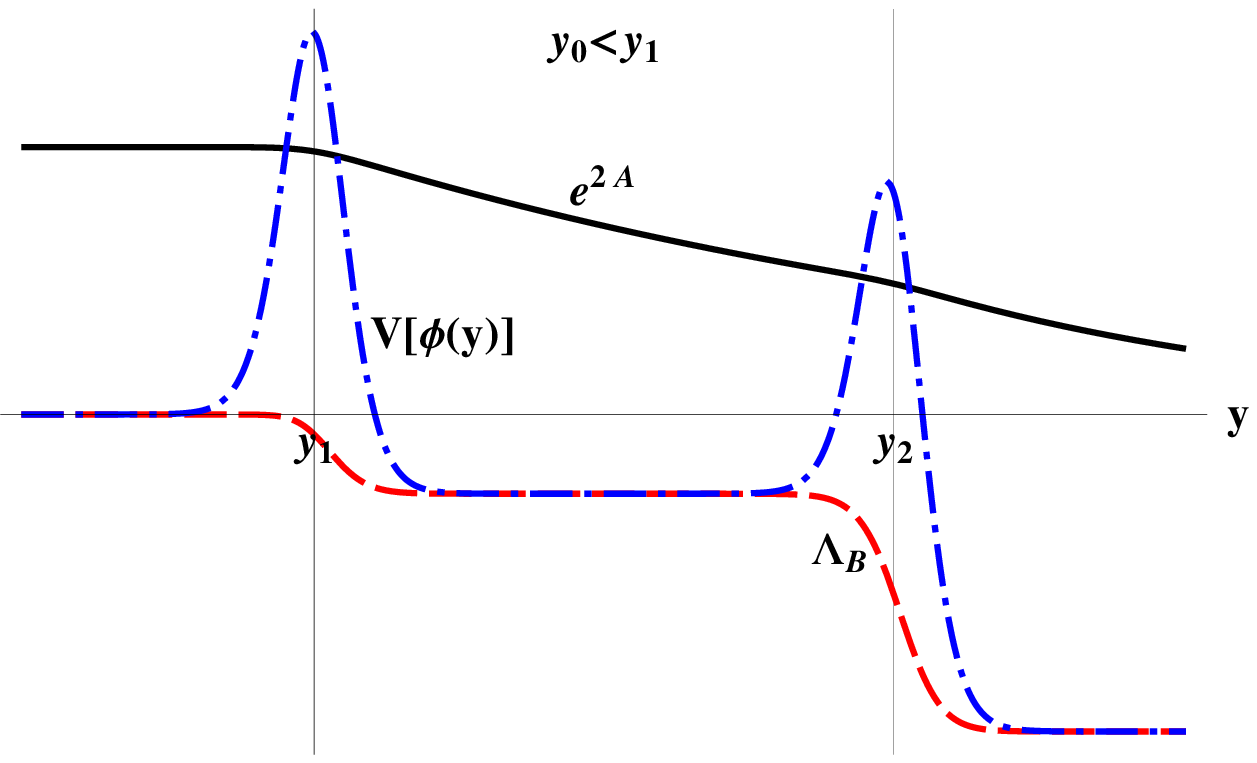}
\end{tabular}
\caption{Graph $(a)$ shows the warped factor $e^{2A}$, the bulk cosmological constant $\Lambda_B$ and
the potential $V[\phi(y)]$ for $y_1<y_0<y_2 $ while graph $(b)$ for $y_0<y_1$. Parameters chosen:
$\beta=3$ and $\kappa_1=\kappa_2=1$.}
\label{fig2}
\end{figure*}

It should be emphasized that the scenario of solving the hierarchy problem described above {\em assumes}
that the Higgs field $H$ could be localized on the IR brane. The issue of localization
is beyond the scope of this paper, however it is being investigated adopting standard
techniques developed so far, see e.g. \cite{Kehagias:2000au,Guo:2011wr,Bajc:1999mh,Liu:2011zy,Liang:2009zzc,Melfo:2006hh,Koley:2004at}, and will be published elsewhere.
\newline\textbf{Case-II:} Lets now consider the case when the extremum position is in between the two
thick branes such that $y_{1}<y_{0}<y_{2}$, in that case
corresponding values of $A^{\prime}(y)$ \eqref{A_prime1} in the brane limit are given by,
\begin{align}
A^{\prime}(y)=\left\{\begin{array}{l c}\frac{1}{12M^{3}} \lambda_{1}\hspace{2cm}& y < y_{1}\\
 0 &y_{1}<y < y_{2}\\
 -\frac{1}{12M^{3}} \lambda_{2} &y > y_{2}\end{array}\right.       \label{Ap2}
\end{align}
This situation is interesting since the 4D graviton is now normalizable and it is localized in between the two positive branes.
However in this scenario the hierarchy problem can not be solved since there is no warping in between the two branes
which is manifest in Fig.~\ref{fig2}-$(a)$. The two positive branes set-up is now similar to the single brane RS-2
\cite{Randall:1999vf}.
\newline\textbf{Case-III:} Now we consider the scenario with extremum located to the left of the left brane or to the right
of the right brane, so  $y_{0}<y_{1}$ or $y_{0}>y_{2}$. For the case when the
extremum lies to the left of $y_1$, in the brane limit, $A^{\prime}(y)$ \eqref{A_prime1} is given by,
\begin{align}
A^{\prime}(y)=\left\{\begin{array}{l c}0\hspace{3cm}& y < y_{1}\\
 -\frac{1}{12M^{3}} \lambda_{1} &y_{1}<y < y_{2}\\
 -\frac{1}{12M^{3}} \left(\lambda_{1}+ \lambda_{2}\right) &y > y_{2}\end{array}\right. ,      \label{Ap1}
\end{align}
In this case we have Minkowski background to left of the brane located at $y_1$ which could be called the UV brane,
however to the right of this brane we have the warped geometry, so that the hierarchy problem could be approached
in the same way as it was discussed for the case-I.
It is worth mentioning that similar geometrical configuration was considered by Gregory, Rubakov, Sibiryakov (GRS) \cite{Gregory:2000jc} with singular branes. The important difference is that GRS model have one positive and one negative tension D3-brane while in our case the both branes are made out of scalar field which mimic two positive tension branes in the brane limit.
The numerical results for corresponding thick branes are shown in Fig.~\ref{fig2}-$(b)$.
We would like to comment here that even though this scenario addresses the hierarchy problem but it does not have the normalizable 4D
graviton, for details see Sec. \ref{Stability of tensor perturbations}. Similarly the other possibility could be considered when the
extremum position is to the right of the brane located at $y_2$, i.e., with $y_{0}>y_{2}$, then in the brane limit the $A^{\prime}(y)$
is given by,
\begin{align}
A^{\prime}(y)=\left\{\begin{array}{l c}\frac{1}{12M^{3}} \left(\lambda_{1}+ \lambda_{2}\right)\hspace{2cm}& y < y_{1}\\
\frac{1}{12M^{3}} \lambda_{2} &y_{1}<y < y_{2}\\
 0 &y > y_{2}\end{array}\right. ,      \label{Ap3}
\end{align}
So in this case geometry to the right of the brane at $y_2$ is Minkowski.

In Table \ref{table_LamB} we summarize results for the cosmological constant in the brane limit,
the regions I, II and III are defined as $y<y_{1}$, $y_{1}<y<y_{2}$ and $y>y_{2}$, respectively.
\begin{table}[h]
  \centering
  \caption{The bulk cosmological constant $\Lambda_{B}$ when $y_{0}\neq y_{1,2}$.}\label{table_LamB}
  \setlength{\tabcolsep}{20pt}
  \begin{tabular}{|c||c|c|c|}
    \hline
     Location of $y_0$ & Region-I & Region-II & Region-III \\ \hline\hline

    $y_{1}<y_{0}<y_{2}$  & $-\frac{1}{6M^{3}}\lambda_{1}^{2}$ & $0$ & $-\frac{1}{6M^{3}}\lambda_{2}^{2}$ \\ \hline
    $y_0<y_{1}$  & $0$ & $-\frac{1}{6M^{3}} \lambda_{1}^{2}$ & $-\frac{1}{6M^{3}} \left(\lambda_{1}+\lambda_{2}\right)^{2}$ \\ \hline
    $y_0>y_{2}$  & $-\frac{1}{6M^{3}}\left(\lambda_{1}+ \lambda_{2}\right)^{2}$ & $-\frac{1}{6M^{3}}\lambda_{2}^{2}$ & $0$ \\
    \hline
  \end{tabular}
\end{table}

In all the above cases, in the brane limit, the $A^{\prime}(y)$ have discontinuities at the brane locations $y=y_{1}$ and $y=y_{2}$.
The discontinuities (or \emph{jump}) are as follows
\begin{align}
[A^{\prime}]_{1}&= -\frac{1}{9M^{3}} \kappa_{1}^{2}= -\frac{1}{12M^{3}}\lambda_{1} \hspace{2cm} y = y_{1}\\
[A^{\prime}]_{2}&= -\frac{1}{9M^{3}} \kappa_{2}^{2}= -\frac{1}{12M^{3}}\lambda_{2} \hspace{2cm} y = y_{2} \label{jump1}
\end{align}
where, $[A^{\prime}]_{i}$ $(i=1,2)$ is defined as,
\[
[A^{\prime}]_{i}\equiv\lim_{\epsilon\to 0}\left[A^{\prime}(y_{i}+\epsilon)-A^{\prime}(y_{i}-\epsilon)\right].
\]
A jump  in $A^{\prime}(y)$ implies that $A^{\prime\prime}(y)$ have delta-like singularity, which is consistent with the equation of
motion \eqref{aeom02}, in fact one can obtain the above jump conditions by integrating Eq.~\eqref{aeom02} from $y_i-\epsilon$ to $y_i+\epsilon$ and
then matching the coefficients of delta functions.

\section{Linearized Einstein equations}
\label{Linearized Einstein equations}
Let us consider fluctuations around the vacuum solution discussed in the previous sections.
We start by replacing the 5D metric $g_{MN}(x,y)$ by $\bar g_{MN}(y)+ h_{MN}(x,y)$, where $\bar g_{MN}(y)$ is
the unperturbed background metric, given as
\begin{align}
\bar g_{\mu\nu}&=e^{2A}\eta_{\mu\nu},\hspace{1cm} \bar g_{\mu5}=0,\hspace{1cm} \bar g_{55}=1. \label{g_MN_per}
\end{align}
It will be convenient to adopt the Einstein equations in the Ricci form as,
\begin{align}
R_{MN}=&\frac{1}{4M^{3}}\tilde{T}_{MN},
\end{align}
where
\begin{align}
\tilde{T}_{MN}&=T_{MN}-\frac{1}{3}g_{MN}T^{A}_{A},\label{bar_T_MN}
\end{align}
where $T_{MN}$ is the energy-momentum tensor given in Eq.~\eqref{emt} while $T^{A}_{A}=-\frac{3}{2}(\nabla\phi)^2-5V(\phi)$ is its trace.
So we get,
\begin{align}
\tilde{T}_{MN}&=\nabla_{M}\phi\nabla_{N}\phi+\frac{2}{3}g_{MN}V(\phi).
\end{align}
The perturbations in the $\tilde{T}_{MN}$ will correspond to fluctuations of the scalar field $\phi(y)$ as $\phi(x,y) = \phi(y)+\varphi(x,y)$ and of the metric $g_{MN}(x,y) = \bar g_{MN}(y)+ h_{MN}(x,y)$. These perturbations
can be calculated order by order in perturbation expansion as,
\[
\tilde{T}_{MN}=\tilde{T}_{MN}^{(0)}+\tilde{T}_{MN}^{(1)}+\cdots,
\]
where ellipses correspond to the higher order fluctuations in $\varphi(x,y)$ and $h_{MN}(x,y)$.
The zeroth and the first order terms are as follows,
\begin{align}
\tilde{T}_{\mu\nu}^{(0)}&=\frac{2}{3}e^{2A}\eta_{\mu\nu}V(\phi), \hspace{1cm} \tilde{T}_{55}^{(0)}=\phi^{\prime2}+\frac{2}{3}V(\phi), \hspace{1cm} \tilde{T}_{\mu5}^{(0)}=0.
\label{T_0}\\
\tilde{T}_{\mu\nu}^{(1)}&=\frac{2}{3}\left(e^{2A}\eta_{\mu\nu}\frac{\partial V(\phi)}{\partial\phi}\varphi+V(\phi)h_{\mu\nu}\right), \label{T_1a}\\
\tilde{T}_{55}^{(1)}&=2\phi^{\prime}\varphi^{\prime}+\frac{2}{3}\frac{\partial V(\phi)}{\partial \phi}\varphi +\frac{2}{3}h_{55}V(\phi), \hspace{1cm} \tilde{T}_{\mu5}^{(1)}=\phi^{\prime}\partial_{\mu}\varphi+\frac{2}{3}h_{\mu5}V(\phi).
\label{T_1b}
\end{align}
Using results from the Appendix~\ref{Conv} one can find the following explicit expressions for
components of the Ricci tensor in the zeroth and first order,
\begin{align}
R_{\mu\nu}^{(0)}=&-e^{2A}\left(A^{\prime\prime}+4A^{\prime2}\right)\eta_{\mu\nu},\hspace{1cm} R_{55}^{(0)}=-4\left(A^{\prime\prime}+A^{\prime2}\right), \hspace{1cm} R^{(0)}_{\mu5}=0, \label{ricci_0}\\
R^{(1)}_{\mu\nu}=&-\frac{1}{2}\partial_{\mu}\partial_{\nu}h_{55}+e^{2A}\eta_{\mu\nu}\left(A^{\prime\prime}+4A^{\prime2}\right)h_{55} +\frac{1}{2}e^{2A}\eta_{\mu\nu}A^{\prime}h_{55}^{\prime}+\frac{1}{2}\left(\partial_{\mu}h_{\nu5}^{\prime}+\partial_{\nu}h_{\mu5}^{\prime}\right)\notag\\
&+A^{\prime}\left(\partial_{\mu}h_{\nu5}+\partial_{\nu}h_{\mu5}\right)-\frac{1}{2}e^{-2A}\Box h_{\mu\nu}+\frac{1}{2}e^{-2A}\eta^{\rho\sigma}\left(\partial_{\mu}\partial_{\rho}h_{\nu\sigma}+\partial_{\nu}\partial_{\rho}h_{\mu\sigma}
-\partial_{\mu}\partial_{\nu}h_{\rho\sigma}\right)\notag\\
&-\frac{1}{2}h_{\mu\nu}^{\prime\prime}-\frac{1}{2}A^{\prime}\eta_{\mu\nu}\eta^{\rho\sigma}h_{\rho\sigma}^{\prime} -A^{\prime2}\left(2h_{\mu\nu}-\eta_{\mu\nu}\eta^{\rho\sigma}h_{\rho\sigma}\right)+A^{\prime}\eta_{\mu\nu}\eta^{\rho\sigma}\partial_{\rho}h_{\sigma5}  , \label{ricci_munu_p}\\
R_{\mu5}^{(1)}=& \frac{1}{2}e^{-2A}\eta^{\rho\sigma}\left(\partial_{\rho}h_{\mu\sigma}^{\prime}-\partial_{\mu}h_{\rho\sigma}^{\prime}\right)
-e^{-2A}A^{\prime}\eta^{\rho\sigma}\left(\partial_{\rho}h_{\mu\sigma}-\partial_{\mu}h_{\rho\sigma}\right) \notag\\
&+\frac{3}{2}A^\prime \partial_{\mu}h_{55}-\frac{1}{2}e^{-2A}\left(\Box h_{\mu5}-\eta^{\rho\sigma}\partial_{\rho}\partial_{\mu}h_{\sigma5}\right)-\left(A^{\prime\prime}+4A^{\prime2}\right)h_{\mu5}, \label{ricci_mu5_p}\\
R_{55}^{(1)}=& e^{-2A}\left(A^{\prime}\eta^{\rho\sigma}h_{\rho\sigma}^{\prime}
+A^{\prime\prime}\eta^{\rho\sigma}h_{\rho\sigma}-\frac{1}{2}\eta^{\rho\sigma}h_{\rho\sigma}^{\prime\prime}\right) -\frac{1}{2}e^{-2A}\Box h_{55} +2A^{\prime}h^{\prime}_{55}+e^{-2A}\eta^{\rho\sigma}\partial_{\rho}h^{\prime}_{\sigma5}, \label{ricci_55_p}
\end{align}
where $\Box$ is the 4D d'Alambertian operator, i.e., $\Box=\eta^{\mu\nu}\partial_{\mu}\partial_{\nu}$.

Having all the components of the Ricci tensor \eqref{ricci_munu_p}-\eqref{ricci_55_p} and $\tilde{T}_{MN}$ \eqref{T_1a}-\eqref{T_1b}, one can write down the equations
of motion for the metric fluctuations $h_{MN}(x,y)$:
\begin{align}
(\mu\nu):\hspace{0.5cm} &-\frac{1}{2}\partial_{\mu}\partial_{\nu}h_{55}+e^{2A}\eta_{\mu\nu}\left(A^{\prime\prime}+4A^{\prime2}\right)h_{55} +\frac{1}{2}e^{2A}\eta_{\mu\nu}A^{\prime}h_{55}^{\prime}+\frac{1}{2}\left(\partial_{\mu}h_{\nu5}^{\prime}+\partial_{\nu}h_{\mu5}^{\prime}\right)\notag\\
&+A^{\prime}\left(\partial_{\mu}h_{\nu5}+\partial_{\nu}h_{\mu5}\right)-\frac{1}{2}e^{-2A}\Box h_{\mu\nu}+\frac{1}{2}e^{-2A}\eta^{\rho\sigma}\left(\partial_{\mu}\partial_{\rho}h_{\nu\sigma}+\partial_{\nu}\partial_{\rho}h_{\mu\sigma}
-\partial_{\mu}\partial_{\nu}h_{\rho\sigma}\right)\notag\\
&-\frac{1}{2}h_{\mu\nu}^{\prime\prime}-\frac{1}{2}A^{\prime}\eta_{\mu\nu}\eta^{\rho\sigma}h_{\rho\sigma}^{\prime} -A^{\prime2}\left(2h_{\mu\nu}-\eta_{\mu\nu}\eta^{\rho\sigma}h_{\rho\sigma}\right)+A^{\prime}\eta_{\mu\nu}\eta^{\rho\sigma}\partial_{\rho}h_{\sigma5}  \notag\\
&=\frac{1}{4M^{3}}\frac{2}{3}\left(e^{2A}\eta_{\mu\nu}\frac{\partial V(\phi)}{\partial\phi}\varphi+V(\phi)h_{\mu\nu}\right), \label{eom_munu}\\
(\mu5):\hspace{0.5cm} &\frac{1}{2}e^{-2A}\eta^{\rho\sigma}\left(\partial_{\rho}h_{\mu\sigma}^{\prime}-\partial_{\mu}h_{\rho\sigma}^{\prime}\right)
-e^{-2A}A^{\prime}\eta^{\rho\sigma}\left(\partial_{\rho}h_{\mu\sigma}-\partial_{\mu}h_{\rho\sigma}\right)+\frac{3}{2}A^\prime \partial_{\mu}h_{55} \notag\\
&-\frac{1}{2}e^{-2A}\left(\Box h_{\mu5}-\eta^{\rho\sigma}\partial_{\rho}\partial_{\mu}h_{\sigma5}\right)=\frac{1}{4M^{3}}\phi^{\prime}\partial_{\mu}\varphi,  \label{eom_mu5}\\
(55):\hspace{0.5cm} &e^{-2A}\left(A^{\prime}\eta^{\rho\sigma}h_{\rho\sigma}^{\prime}
+A^{\prime\prime}\eta^{\rho\sigma}h_{\rho\sigma}-\frac{1}{2}\eta^{\rho\sigma}h_{\rho\sigma}^{\prime\prime}\right) -\frac{1}{2}e^{-2A}\Box h_{55} +2A^{\prime}h^{\prime}_{55}+e^{-2A}\eta^{\rho\sigma}\partial_{\rho}h^{\prime}_{\sigma5}\notag\\
&=\frac{1}{4M^{3}}\left(2\phi^{\prime}\varphi^{\prime}+\frac{2}{3}\frac{\partial V(\phi)}{\partial \phi}\varphi +\frac{2}{3}h_{55}V(\phi)\right).\label{eom_55}
\end{align}

Additionally, we also have the equation of motion of the scalar field $\phi$ \eqref{eineq2} in the first order in the fluctuations $h_{MN}(x,y)$ and $\varphi(x,y)$ as,
\begin{align}
e^{-2A}\Box\varphi+\varphi^{\prime\prime}+4A^{\prime}\varphi^{\prime}-\frac{\partial^{2}V(\phi)}{\partial\phi^{2}}\varphi
+\frac{1}{2}\phi^{\prime}\left(e^{-2A}h\right)^{\prime}-\frac{1}{2}\phi^{\prime}h^{5\prime}_{5}-\left(\phi^{\prime\prime}+4A^{\prime}\phi^{\prime}\right)h_5^5&=0,
\label{phi_cons_1}
\end{align}
where $h\equiv\eta^{\mu\nu}h_{\mu\nu}$.

In the remaining part of this section we will derive equations of motion for perturbations of the metric and the scalar field. We are going to adopt a decomposition of the metric perturbation $h_{MN}$ into \emph{scalar}, \emph{vector} and \emph{tensor} (SVT) components. For completeness we review SVT perturbations in the Appendix~\ref{SVT decomposition of perturbations and gauge choice} where it is shown by \eqref{pert_munu_s}-\eqref{pert_55_s} that the SVT modes decouple, therefore in the following subsections we discuss them separately case-by-case.

\subsection{Scalar perturbations}
\label{Scalar perturbations}
Scalar perturbations contribute to the metric as follows
\begin{align}
ds^{2}&=e^{2A}\left[\left(1-2\psi\right)\eta_{\mu\nu}-2\partial_{\mu}\partial_{\nu}E \right]dx^{\mu}dx^{\nu}
+\partial_{\mu}Bdx^{\mu}dy+\left(1+2\chi\right)dy^{2},
\label{metric_per}
\end{align}
The scalar modes appearing here are not gauge invariant, i.e., their values are affected by the choice of different
coordinates. It is therefore instructive to either work with the gauge invariant quantities or choose a suitable
gauge such that the ambiguities related to the coordinate transformations can be removed. Here we choose the
longitudinal gauge such that the gauge freedom is fixed
completely, as discussed in Appendix~\ref{SVT decomposition of perturbations and gauge choice}. Therefore, for the scalar modes of perturbation, we have $B=E=0$~\footnote{We suppress
the $\check {\phantom{B}} $ signs hereafter, as it is clear that we are referring the modes in the new reference frame as discussed in Appendix~\ref{SVT decomposition of perturbations and gauge choice}.}.

\noi In the longitudinal gauge the perturbed metric \eqref{metric_per} is of the form,
\begin{align}
ds^{2}&=e^{2A}\left(1-2\psi\right)\eta_{\mu\nu}dx^{\mu}dx^{\nu}+\left(1+2\chi\right)dy^{2}.
\label{metric_Newton}
\end{align}
Adopting the general results from the Appendix~\ref{SVT decomposition of perturbations and gauge choice}
we find the following form of the linearized field equations for the scalar modes,
\begin{align}
(\mu\nu):\hspace{1cm} &e^{2A}\eta_{\mu\nu}\bigg[2\left(A^{\prime\prime}+4A^{\prime2}\right)\chi+A^{\prime}\chi^{\prime} +e^{-2A}\Box\psi+8A^{\prime}\psi^{\prime}+\psi^{\prime\prime}\bigg]\notag\\
& \hspace{4cm}+ \partial_{\mu}\partial_{\nu}\left(2\psi-\chi\right)= \frac{1}{6M^{3}}e^{2A}\eta_{\mu\nu}\frac{\partial V(\phi)}{\partial \phi}\varphi,
\label{spert_munu}\\
(\mu5):\hspace{1cm} &3A^{\prime}\partial_{\mu}\chi+3\partial_{\mu}\psi^{\prime}=\frac{1}{4M^{3}}\phi^{\prime}\partial_{\mu}\varphi,
\label{spert_mu5}\\
(55):\hspace{1cm} &4\left(\psi^{\prime\prime}+2A^{\prime}\psi^{\prime}\right)+4A^{\prime}\chi^{\prime}-e^{-2A}\Box \chi =\frac{1}{4M^{3}}\left[2\phi^{\prime}\varphi^\prime+\frac{2}{3}\frac{\partial V(\phi)}{\partial \phi}\varphi +\frac{4}{3}V(\phi)\chi\right].
\label{spert_55}
\end{align}
One can notice from Eq.~\eqref{spert_munu} that the absence of the $\partial_{\mu}\partial_{\nu}$ term on the right hand side implies $\partial_{\mu}\partial_{\nu}\left(2\psi-\chi\right)=0$ so that $\chi=2\psi+c(y)$. Where $c(y)$ is a $y$-dependent constant of integration which can be fixed by the requirement that at 4D infinities $\chi, \psi \to 0$, therefore $c(y)=0$.

When $\chi=2\psi$ the equation of motion for the scalar field fluctuation \eqref{phi_cons_1} simplifies,
\begin{align}
e^{-2A}\Box\varphi+\varphi^{\prime\prime}+4A^{\prime}\varphi^{\prime}-\frac{\partial^{2}V(\phi)}{\partial\phi^{2}}\varphi
-6\phi^{\prime}\psi^{\prime}-4 \left(\phi^{\prime\prime}+4A^{\prime}\phi^{\prime}\right)\psi&=0.
\label{sphi_cons}
\end{align}
It is important to note that, as usually in such cases, the equations of motions \eqref{spert_munu}-\eqref{sphi_cons}
are not independent. Adopting the relation $\chi=2\psi$ and the background equations of motion
one derive the following equation that we will use instead of \eqref{spert_munu} and \eqref{spert_55}
\begin{align}
3\psi^{\prime\prime}+6A^{\prime}\psi^{\prime}-3e^{-2A}\Box\psi =\frac{1}{2M^{3}}\phi^{\prime}\varphi^\prime.
\label{spsi_sub}
\end{align}
Hence, Eqs. \eqref{spert_mu5}, \eqref{sphi_cons} and \eqref{spsi_sub} complete the set of linearized
equations for the scalar modes. In the subsequent section we will use these equations to study stability
of scalar field perturbations.

\subsection{Vector perturbations}
\label{Vector perturbations}
We can write down the metric for the vector perturbations as,
\begin{align}
ds^{2}&=e^{2A}\left(\eta_{\mu\nu}+\partial_{\mu}G_{\nu}+\partial_{\nu}G_{\mu}\right)dx^{\mu}dx^{\nu}
+C_{\mu}dx^{\mu}dy+dy^{2}, \label{metric_V_per}
\end{align}
where $C_{\mu}$ and $G_{\mu}$ are divergenceless vectors defined in Eqs. \eqref{h_munu} and \eqref{h_mu5}.
Adopting the general results from the Appendix~\ref{SVT decomposition of perturbations and gauge choice}
we find the following form of the linearized field equations for the vector modes
\begin{align}
\partial_{\mu}\bigg[C_{\nu}^{\prime}+2A^{\prime}C_{\nu}+
-4e^{2A}A^{\prime}G_{\nu}^{\prime}-e^{2A}G_{\nu}^{\prime\prime}\bigg]&=0,
\label{vector_p1}\\
e^{-2A}\Box C_{\nu} - \Box G_\mu^\prime &= 0.
\label{vector_p2}
\end{align}
Since we are working in the gauge where $G_\mu=0$ so the equations of motion for the vector modes of the metric
perturbations read
\begin{align}
\Box C_{\nu}&=0, \hspace{1cm} \partial_{\mu}\left(C_{\nu}^{\prime}+2A^{\prime}C_{\nu}\right)=0.
\label{vector_p5}
\end{align}

\subsection{Tensor perturbations}
\label{Tensor perturbations}
The tensor metric perturbation \eqref{metric} can be written as,
\begin{align}
ds^{2}&=e^{2A(y)}(\eta_{\mu\nu}+H_{\mu\nu})dx^{\mu}dx^{\nu}+dy^{2},
\label{metricT}
\end{align}
where, $H_{\mu\nu}=H_{\mu\nu}(x,y)$ is the tensor fluctuation as defined in \eqref{h_munu}.
Adopting the general results from the Appendix~\ref{SVT decomposition of perturbations and gauge choice}
we find the following form of the linearized field equations for the tensor modes
\begin{align}
\left(\partial_{5}^{2}+4A^{\prime}\partial_{5}+e^{-2A}\Box\right)H_{\mu\nu} &=0.
\label{graviton_0}
\end{align}
The zero-mode solution (corresponding to $\Box H_{\mu\nu}=0$) of the above equation should represent
the 4D graviton while the non-zero modes are the Kaluza-Klein (KK) graviton excitations.

The issue of the stability of the background solution (in terms of the SVT components)
will be discussed in the next section.

\section{Stability of the solutions}
\label{Stability of the SVT perturbations}

\subsection{Tensor perturbations}
\label{Stability of tensor perturbations}
\begin{figure*}[ht]
\begin{tabular}{cc}
($a$)&($b$)\\
\includegraphics[scale=0.55]{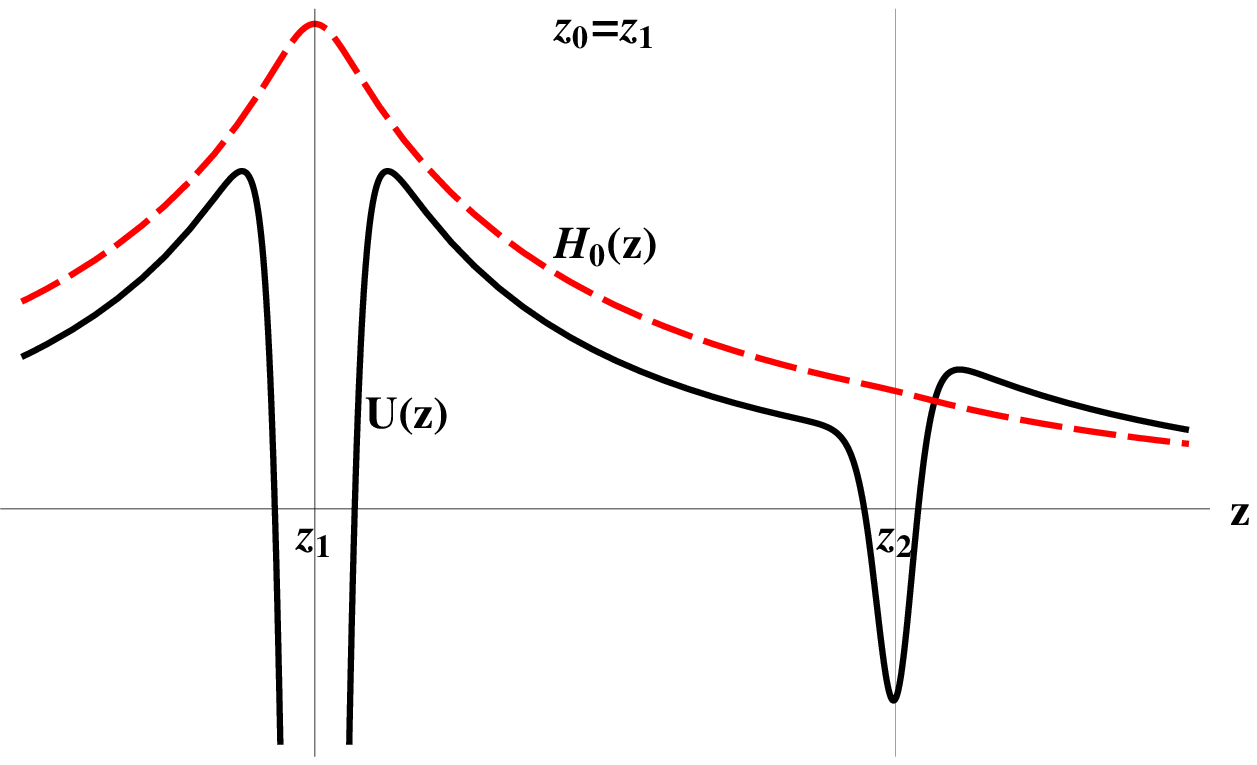} \
& \ \includegraphics[scale=0.55]{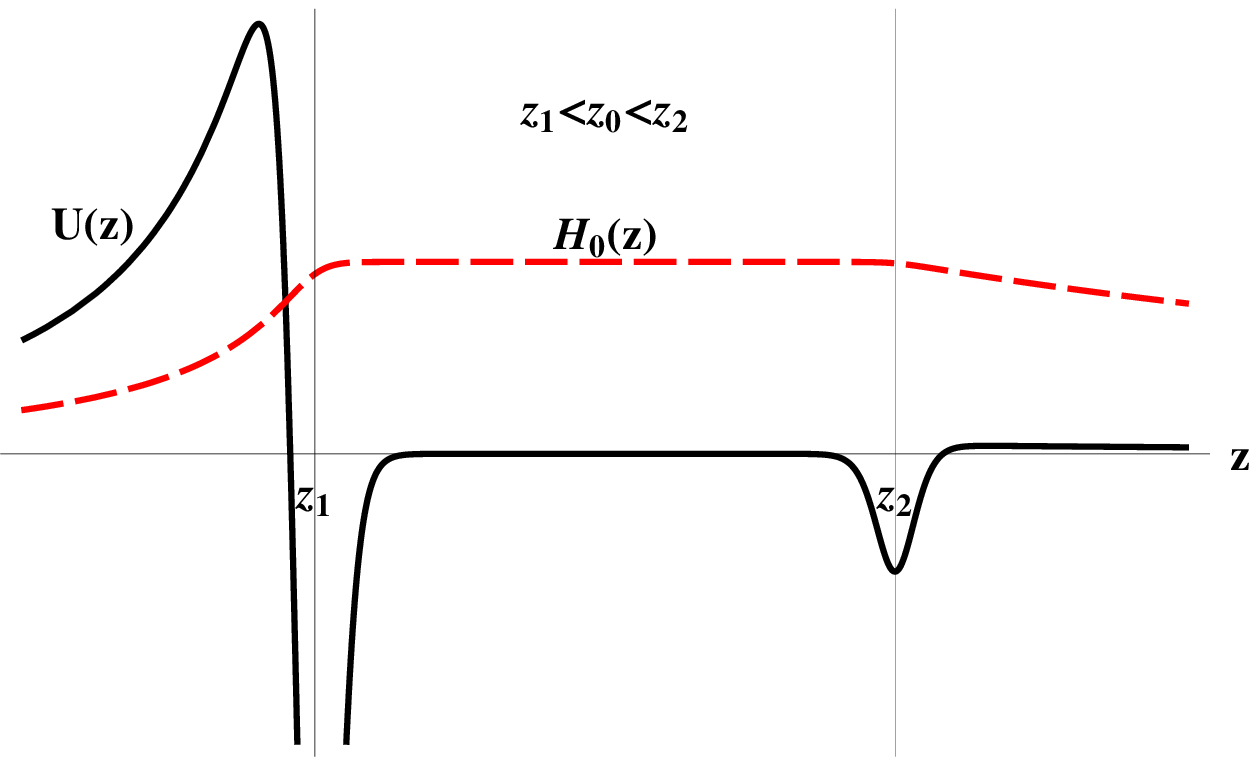}\\
($c$)&($d$)\\
\includegraphics[scale=0.55]{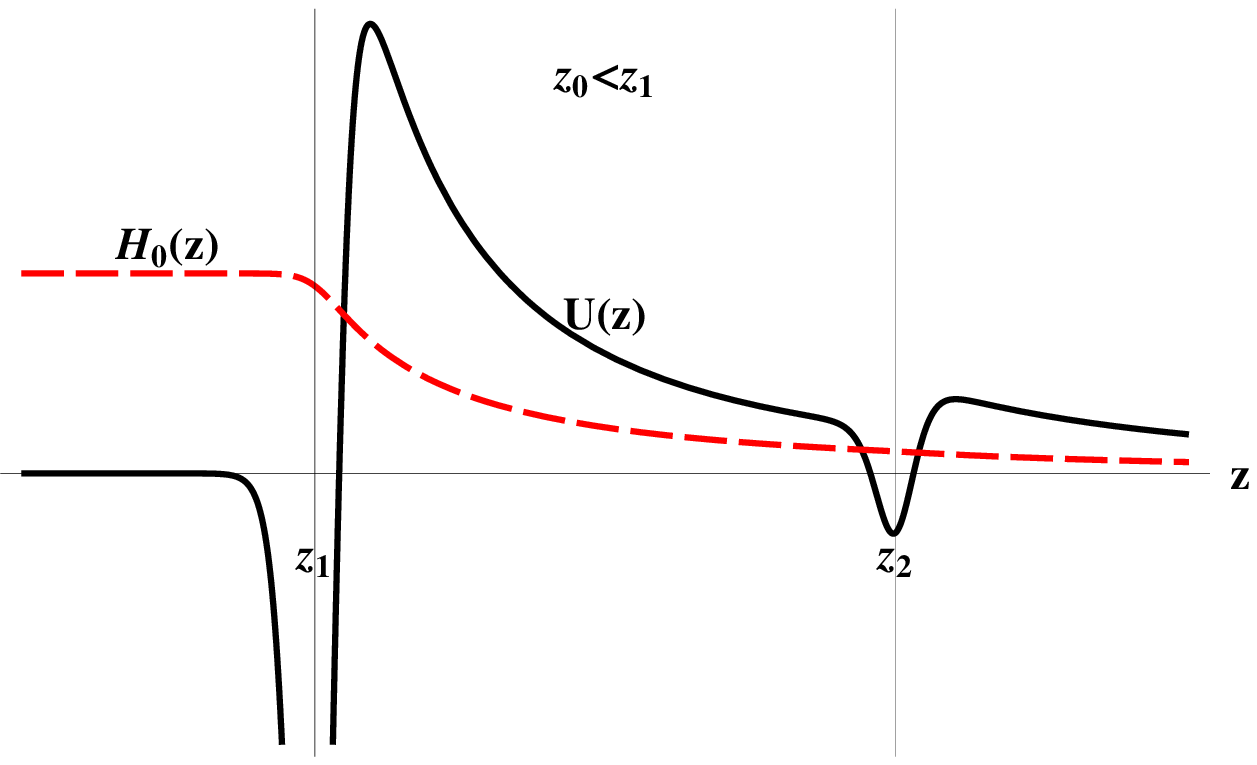} \
& \  \includegraphics[scale=0.55]{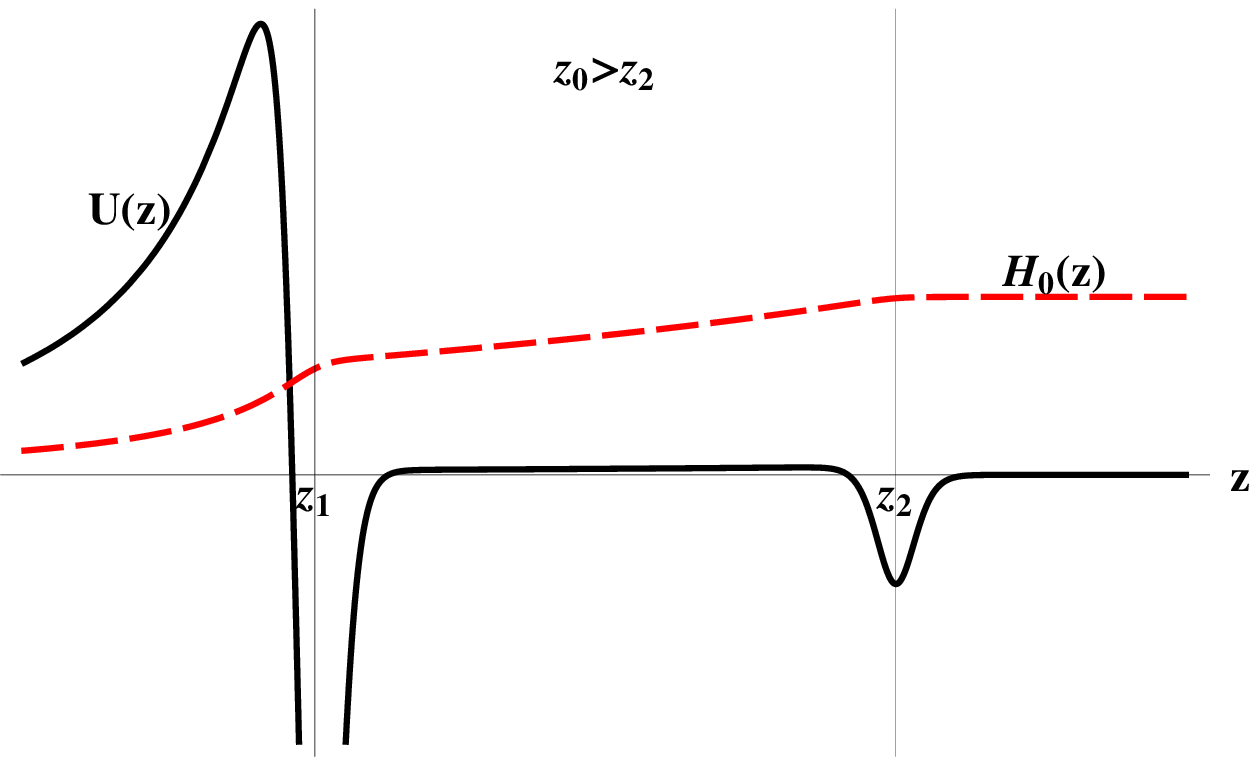}
\end{tabular}
\caption{These graphs illustrate the shape of the quantum mechanics potential $U(z)$ in solid (black) for all the for scenarios that we have considered in Sec. \ref{The brane limit} and the corresponding shape of the zero-mode (4D graviton) in dashed (red) curve. Parameters chosen: $\beta=5$, $\kappa_1=3$ and $\kappa_2=1$.}
\label{fig3}
\end{figure*}

In order to gain more intuition and understanding of the graviton equation of motion (\ref{graviton_0}),
it is convenient to change
the variables such that we get rid of the exponential factor in front of the d'Alambertian and the single derivative term with
$A^{\prime}$, so that we convert the above equation into the standard Schr\"odinger like form. We can achieve this in two
steps; first by changing coordinates such that the metric becomes conformally flat:
\begin{align}
ds^{2}&=e^{2A(z)}\left(g_{\mu\nu}dx^{\mu}dx^{\nu}+dz^{2}\right),
\label{metric1}
\end{align}
where $dy=e^Adz$. In the new coordinates the Eq.~\eqref{graviton_0} takes the form
\begin{align}
\left(\partial_{z}^{2}+3\dot{A}(z)\partial_{z}+\Box\right)H_{\mu\nu} &=0, \label{graviton_1}
\end{align}
where $dot$ over $A$ represent the derivative with respect to $z$ coordinate. Now we can perform the second step
removing the single derivative term in \eqref{graviton_1} by making the following
redefinition of the graviton field
\begin{align}
H_{\mu\nu}(x,z) &=e^{-3A/2}\tilde{H}_{\mu\nu}(x,z).
\label{hat_h0}
\end{align}
So the Eq.~\eqref{graviton_1} will take the form of the Schr\"odinger equation,
\begin{align}
\left(\partial_{z}^{2}-\frac{9}{4}\dot{A}^{2}(z)-\frac{3}{2}\ddot{A}(z)+\Box\right)\tilde{H}_{\mu\nu}(x,z) &=0.
\label{schrodinger_eq_0}
\end{align}
We can further split the $\tilde H_{\mu\nu}(x,z)$ into $\tilde H_{\mu\nu}(x,z)=\hat H_{\mu\nu}(x)\bar{H}(z)$, where $\hat H_{\mu\nu}(x)=e^{ipx}$ is a $z$-independent plane wave such that $\Box \hat H_{\mu\nu}(x)=m^2\hat H_{\mu\nu}(x)$, with $-p^2=m^2$ being the 4D KK mass of the fluctuation. Then the above equation takes the form,
\begin{align}
\left(\partial_{z}^{2}-\frac{9}{4}\dot{A}^{2}(z)-\frac{3}{2}\ddot{A}(z)+m^{2}\right)\bar{H}(z) &=0,\\
\left[-\partial_{z}^{2}+U(z)\right]\bar{H}(z) &=m^{2}\bar{H}(z), \label{schrodinger_eq}
\end{align}
where $U(z)$ is the potential,
\begin{align}
U(z) &=\frac{9}{4}\dot{A}^{2}(z)+\frac{3}{2}\ddot{A}(z).
\label{U_potential}
\end{align}
Note that we can write the Schr\"odinger-like equation \eqref{schrodinger_eq} in supersymmetric quantum mechanics form as,
\beq
{\cal Q}^{\dagger}{\cal Q}\bar H  =\left(-\partial_{z}-\frac{3}{2}\dot{A}\right)\left(\partial_{z}-\frac{3}{2}\dot{A}\right)\bar{H}
=m^{2}\bar{H}.
\label{Susy_QM}
\eeq
The zero mode ($m^2=0$) profile, $\bar H_0(z)$, corresponds to the graviton in the 4D effective theory. The stability
with respect to the tensor fluctuations of the background solution is guaranteed by the positivity of the operator
${\cal Q}^{\dagger}{\cal Q}$ in the supersymmetric version of the equation of motion \eqref{Susy_QM} as it forbids the existence of tachyonic modes with negative mass$^2$, $m^{2}<0$~\footnote{Since $\int dz ({\cal Q} \bar{H})^2+\bar{H} {\cal Q} \bar{H}\left|_{-\infty}^{+\infty}\right. = m^2 \int dz \bar{H}^2$, therefore in order to guarantee $m^2>0$ the
boundary term must vanish or be positive. Note that for the zero mode ${\cal Q} \bar{H}_0=0$, so indeed the boundary term disappears. }.
So, in that case, the perturbation is not growing in time. $\bar{H}_0(z)$ could be
found by noticing that the annihilation operator ${\cal Q}$ should vanish acting on $\bar H_0$,
\begin{align}
{\cal Q}\bar H_0 &=\left(\partial_{z}-\frac{3}{2}\dot{A}\right)\bar{H}_0=0,
\label{Susy_QM_0}
\end{align}
which implies that,
\begin{align}
\bar H_0(z) &=e^{\frac{3}{2}A(z)} .
\label{bar_H_0}
\end{align}
For massive KK modes we need to solve the Eq.~\eqref{schrodinger_eq} with $m^2\neq 0$.
For large $z$ the potential $U(z)$ goes to zero for the case $(i)$ and $(ii)$ as shown in
Fig.~\ref{fig3}$(a)$ and $(b)$, so
Eq.~\eqref{schrodinger_eq} reduces to one dimensional Klein-Gordon (KG) equation, i.e.,
\begin{align}
\left(\partial_{z}^{2}+m^2\right)\bar{H}_m(z) &=0. \label{KG_eq}
\end{align}
Therefore in the large $z$ limit, we expect,
\begin{align}
\bar{H}_m(z) &\approx c_1\cos(mz)+c_2\sin(mz),
\label{KK_mode}
\end{align}
where $c_1$ and $c_2$ are constants. Therefore the massive KK modes are plane wave normalizable and
we have a continuum spectrum of KK states for the cases $(i)$ and $(ii)$ discussed in Sec. \ref{The brane limit}.
Before closing this subsection we will briefly comment on the effective 4D gravity.
We are going to estimate the effective 4D Plank mass and discuss the localization of the
zero-mode of the perturbation and then corrections to the Newton's potential due to the
massive KK modes. To calculate the 4D Plank mass it is important to note that
Eq.~\eqref{schrodinger_eq_0} only
involves 2nd derivatives of the metric perturbation $\tilde{H}_{\mu\nu}(x,z)$ which is related to
the fact that in the action these fluctuations have the following canonical kinetic term
\begin{align}
S\approx M^{3}\int d^{4}x dz \partial_{M}\tilde{H}_{\mu\nu}(x,z)\partial^{M}\tilde{H}^{\mu\nu}(x,z)+
\cdots,
\label{action_KT_1}
\end{align}
where the indices are contracted with the 5D Minkowski metric $\eta_{MN}$. Since we have $\tilde{H}_{\mu\nu}(x,z)=\hat{H}_{\mu\nu}(x)\bar{H}(z)$ therefore \eqref{action_KT_1} could be
rewritten as follows
\begin{align}
S\approx M^{3}\int dz \bar H^{2}(z) \int d^{4}x\partial_{\alpha}\hat{H}_{\mu\nu}(x)\partial^{\alpha}\hat{H}^{\mu\nu}(x)+
\cdots, \label{action_KT_2}
\end{align}
from this we can read out the effective 4D linearized gravity as,
\begin{align}
S\approx M^{2}_{Pl} \int d^{4}x\partial_{\alpha}\hat{H}_{\mu\nu}(x)\partial^{\alpha}\hat{H}^{\mu\nu}(x)+\cdots,
\label{action_4D}
\end{align}
where $M_{Pl}$ is the effective 4D Planck mass, i.e.,
\begin{align}
M^{2}_{Pl}&= M^{3}\int dz \bar H^{2}(z),
\label{4D_Planck mass}
\end{align}
where $\bar H(z)$ satisfies the supersymmetric quantum mechanic equation \eqref{Susy_QM} for $m^2=0$.
In order to reproduce the standard 4D General Relativity, $M^{2}_{Pl}$ must be finite, in other words
$\bar H_0(z)$ must be normalizable.
It is easy to see from (\ref{bar_H_0}) that indeed $\bar H_0(z)$ is normalizable for the cases
$(i)$ and $(ii)$ for which the warp function $A(y)$ posses the following asymptotic
behavior (see Eqs. \eqref{Ap4} and \eqref{Ap3}),
\begin{align}
A^\prime(y) &<0 \hspace{0.5cm} \text{as} \hspace{0.5cm} y\to\infty, \label{A_prime_norm_1}\\
A^\prime(y) &>0 \hspace{0.5cm} \text{as} \hspace{0.5cm} y\to-\infty. \label{A_prime_norm_2}.
\end{align}
The above implies that
\begin{align}
\int dz \bar H_0^{2}(z) = \int dz e^{3A(z)} = \int dy e^{2A(y)} < \infty,
\label{normalizablity_y}
\end{align}
therefore $\bar H_0(z)$ is normalizable (see also Fig.~\ref{fig3} $(a)$ and $(b)$) and
$M^{2}_{Pl}$ is finite for the cases $(i)$ and $(ii)$. The situation for the case $(iii)$ is
far more complicated, as there neither we have the finite 4D effect Planck mass nor we have a
normalizable zero-mode (see Fig.~\ref{fig3} $(c)$ and $(d)$). However, as it was pointed out for
similar singular brane set-up (GRS \cite{Gregory:2000jc}), the effective 4D gravity
on the brane can be reproduced and we could have the quasi-localized gravity \cite{Csaki:2000ei,Csaki:2000pp,Gregory:2000iu,Dvali:2000rv}. We are not going to discuss that
case any farther.

In order to build some intuition concerning possible size of corrections to the Newton's
law due to an exchange of massive KK modes, we will briefly consider a simple case with two point
like sources $m_1$ and $m_2$ located on the 4D slice at $z=z_1=0$ and at $z=z_2$ in the 5th dimension,
so at locations of perspective branes in the brane limit. With these assumptions, the corrections to
the Newton's law for the case ($i$) (the thick brane version of Lykken-Randall model)
could be easily obtained. At the UV brane ($z_1=0$) the potential (after integrating over
graviton KK modes) is modified as follows
\begin{align}
{\cal U}_{UV}(z) &\approx G_N\frac{m_1 m_2}{r}+\frac{1}{M^3}\int_0^\infty dm\frac{m_1 m_2e^{-mr}}{r}\bar H_m(0)^2,
\label{Newton_law_UV}
\end{align}
where $G_N=(8\pi M_{Pl}^2)^{-1}$ is the 4D Newton's constant, $\bar H_m(0)$ is the value of the
graviton wave function at the UV brane and $r$ is the distance between the point sources on the
4D slice located at $z=z_1=0$. Now, if sources are located at $z=z_2$
the gravitational potential is modified as follows
\begin{align}
{\cal U}_{IR}(z) &\approx G_N\frac{m_1 m_2}{r}+
\frac{1}{M^3}\int_0^\infty dm\frac{m_1 m_2e^{-mr}}{r}e^{3A(z_2)}\bar H_m(z_2)^2,
\label{Newton_law_IR}
\end{align}
where $\bar H_m(z_2)$ is the value of the graviton wave function at the IR brane.
For singular branes the exact analysis performed in various set-ups show \cite{Randall:1999vf,Lykken:1999nb,Gregory:2000jc,Csaki:2000ei,Csaki:2000zn,Csaki:2000pp,Gregory:2000iu,Dvali:2000rv}
that corrections to the Newton's law are small and the usual 4D gravity is restored on IR
branes, therefore one could expect that similar conclusions hold also for our thick brane scenarios.
In fact, as it was pointed out by Csaki et al. \cite{Csaki:2000fc} also in thick brane scenario
corrections due to massive KK modes do not induce any harmful effects for the 4D
effective gravity. The detailed and rigorous study of these issues for our two thick branes,
is beyond the scope of this paper and will be considered somewhere else.

\subsection{Scalar perturbations}
\label{Stability of scalar perturbations}
\begin{figure*}[ht]
\begin{tabular}{cc}
($a$)&($b$)\\
\includegraphics[scale=0.55]{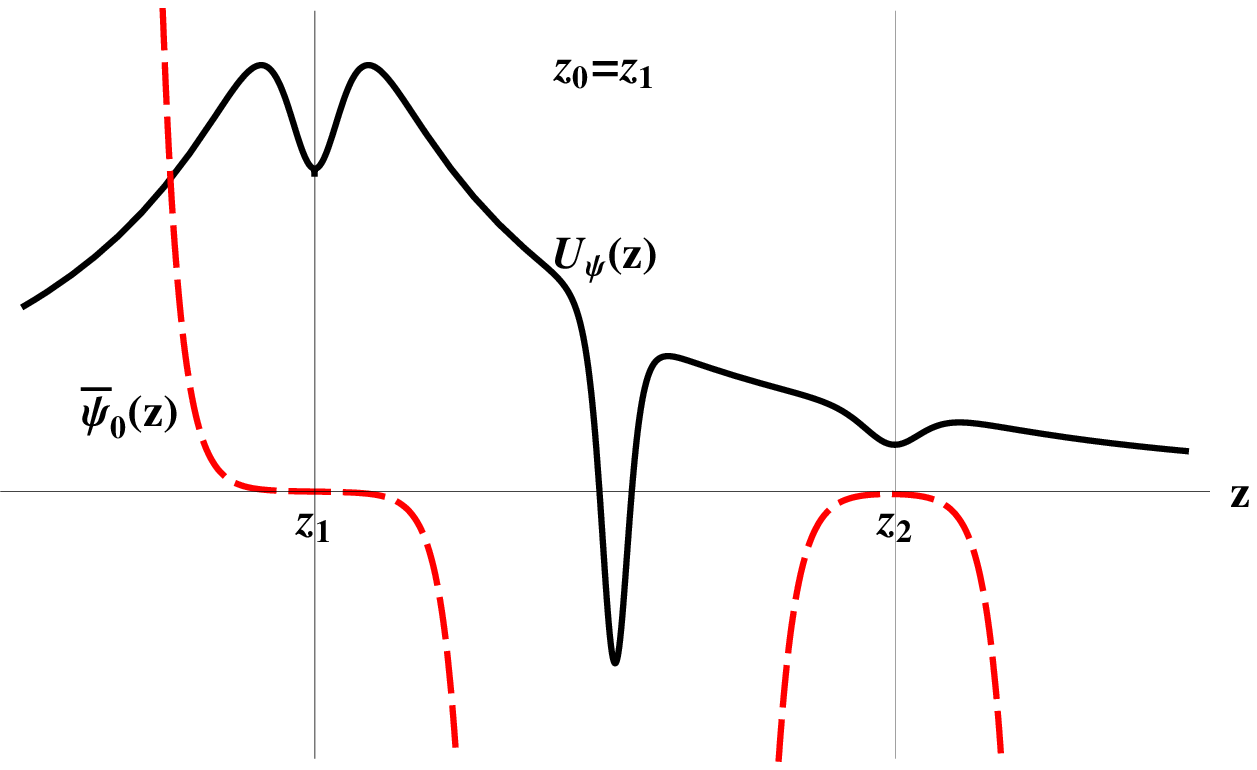} \
& \ \includegraphics[scale=0.55]{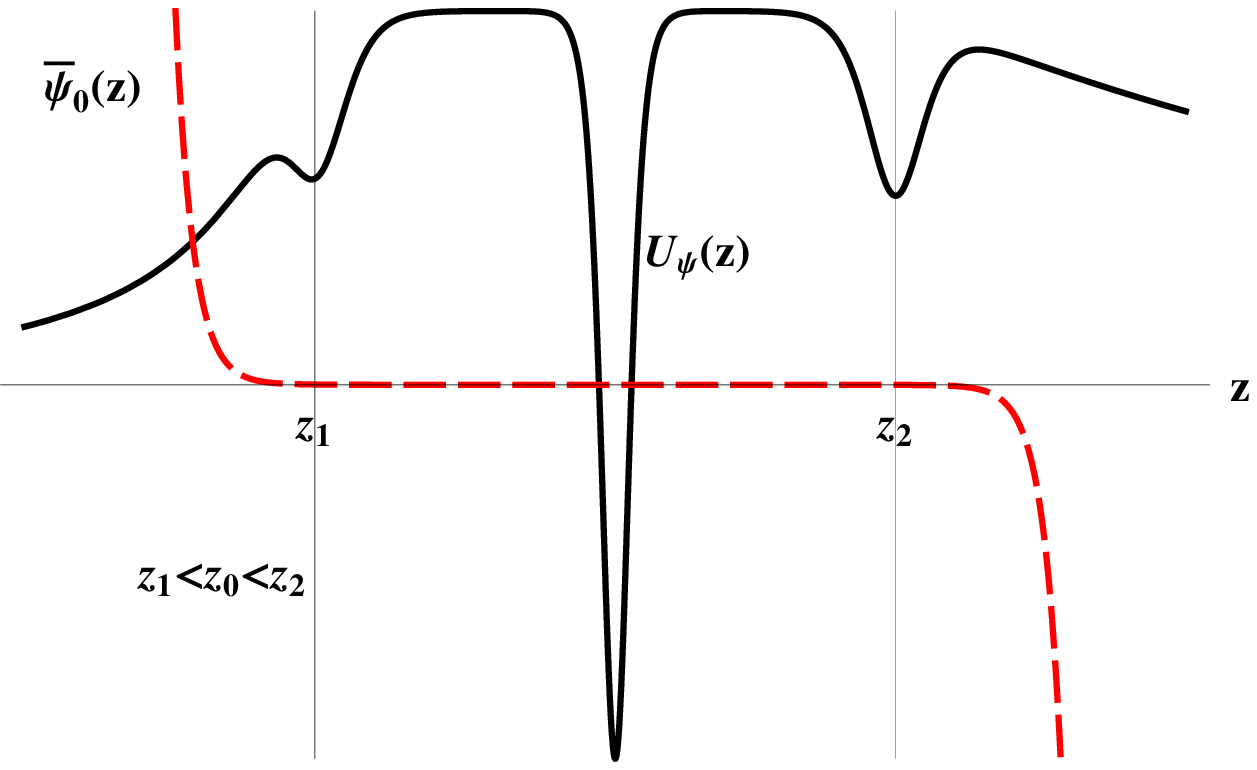}\\
($c$)&($d$)\\
\includegraphics[scale=0.55]{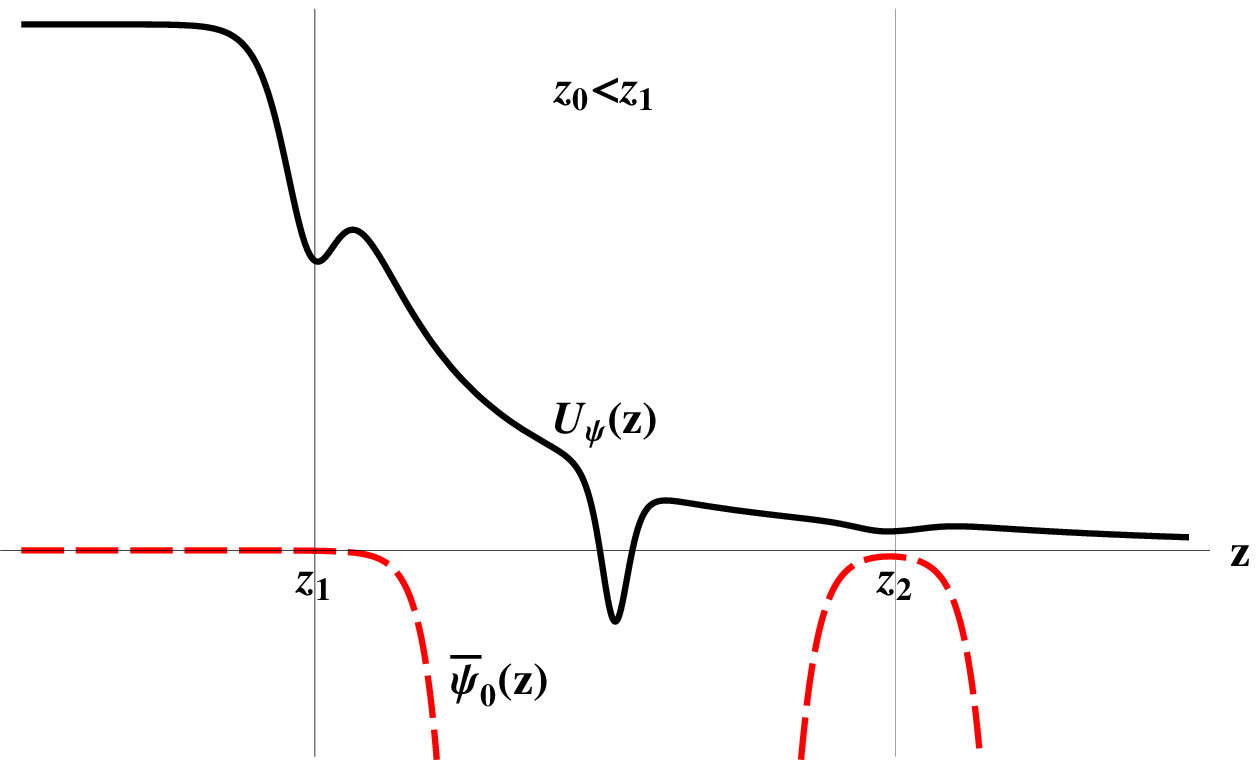} \
& \  \includegraphics[scale=0.55]{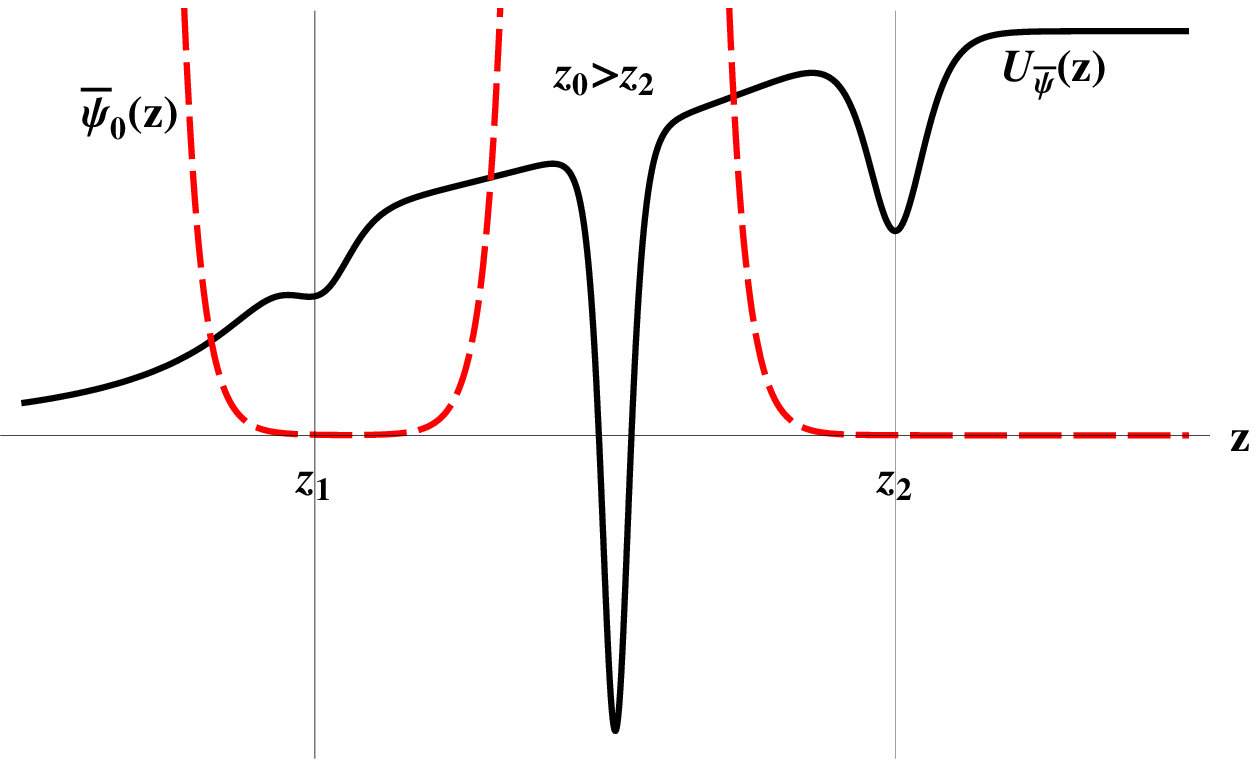}
\end{tabular}
\caption{These graphs illustrate the shape of the quantum mechanics potential $U_{\bar\psi}(z)$ \eqref{scalar_potential}
in solid (black) for all the for scenarios that we have considered in Sec. \ref{The brane limit} and the
corresponding shape of the scalar zero-mode in dashed (red) curve. The parameters are same as in Fig. \ref{fig3}.}
\label{fig4}
\end{figure*}

The linearized field equations corresponding to the scalar modes of the perturbation are given by
Eqs. \eqref{spert_mu5}, \eqref{sphi_cons} and \eqref{spsi_sub}, i.e.,
\begin{align}
6A^{\prime}\partial_{\mu}\psi+3\partial_{\mu}\psi^{\prime}&=\frac{1}{4M^{3}}\phi^{\prime}\partial_{\mu}\varphi,
\label{scalar_pert_1}\\
e^{-2A}\Box\varphi+\varphi^{\prime\prime}+4A^{\prime}\varphi^{\prime}-\frac{\partial^{2}V(\phi)}{\partial\phi^{2}}\varphi
&-6\phi^{\prime}\psi^{\prime}-4\frac{\partial V(\phi)}{\partial\phi}\psi=0,
\label{scalar_pert_3}\\
\psi^{\prime\prime}+2A^{\prime}\psi^{\prime}-e^{-2A}\Box\psi &=\frac{1}{6M^{3}}\phi^{\prime}\varphi^\prime.
\label{scalar_pert_2}
\end{align}
One can integrate Eq.~\eqref{scalar_pert_1} over $x$-coordinates and get the following equation,
\begin{align}
6A^{\prime}\psi+3\psi^{\prime}&=\frac{1}{4M^{3}}\phi^{\prime}\varphi,
\label{scalar_pert_01}
\end{align}
where we have put the $y$-dependent integration constant to zero by the requirement that the perturbations
vanish at 4D infinities.
It is more convenient to use the conformal frame where the metric can be written
as in Eq.~\eqref{metric1} such that $dy=e^{A(y)} dz$. Hence, in the new coordinates our equations of motion
\eqref{scalar_pert_3}-\eqref{scalar_pert_01} take the following form,
\begin{align}
\Box\varphi+\ddot\varphi+3\dot A\dot\varphi-e^{2A}\frac{\partial^{2}V(\phi)}{\partial\phi^{2}}\varphi
&-6\dot\phi\dot\psi-4e^{2A}\frac{\partial V(\phi)}{\partial\phi}\psi=0,
\label{scalar_pert_3b}\\
\ddot{\psi}+\dot A\dot{\psi}-\Box\psi&=\frac{1}{6M^{3}}\dot\phi\dot{\varphi},
\label{scalar_pert_2b}\\
2\dot A\psi+\dot{\psi}&=\frac{1}{12M^{3}}\dot\phi \varphi,
\label{scalar_pert_1b}
\end{align}
First we solve Eq.~\eqref{scalar_pert_1b} with respect to $\varphi$ and calculate $\dot\varphi$ as,
\begin{align}
\dot\varphi&=\frac{12M^{3}}{\dot\phi^2}\left[\left(2\ddot A\psi+2\dot A\dot\psi+\ddot\psi\right)\dot\phi-
\left(2\dot A\psi+\dot{\psi}\right)\ddot\phi\right] , \label{dot_varphi}
\end{align}
and then use it in \eqref{scalar_pert_2b}, so that we obtain an equation only for $\psi$,
\begin{align}
\ddot{\psi}+\left(3\dot A-2\frac{\ddot\phi}{\dot\phi}\right)\dot{\psi}+\left(4\ddot A -4\dot A \frac{\ddot\phi}{\dot\phi}+\Box\right)\psi&=0.
\label{scalar_pert_2c}
\end{align}
To convert this equation into the Schr\"odinger form it is instructive to remove the first derivative terms of the
perturbation $\psi$, to do so we redefine the scalar perturbations as,
\begin{align}
\psi(x,z)=e^{-\frac{3}{2}A(z)}\dot\phi\tilde\psi(x,z).
\label{tilde_psi}
\end{align}
Then the linearized field equation for the scalar perturbation $\psi$ takes the following form,
\begin{align}
-\ddot{\tilde\psi}+\left[\frac{9}{4}\dot A^2 -\frac{5}{2}\ddot A+\dot A \frac{\ddot\phi}{\dot\phi}+
2\left(\frac{\ddot\phi}{\dot\phi}\right)^2-\frac{\dddot\phi}{\dot\phi}\right]\tilde\psi&=\Box\tilde\psi.
\label{scalar_pert_2d}
\end{align}
We can further decompose the $\tilde \psi(x,z)$ into $\tilde \psi(x,z)=\hat \psi(x)\bar{\psi}(z)$, where $\hat \psi(x)=e^{ipx}$
is a $z$-independent plane wave such that $\Box \hat \psi(x)=m^2\hat \psi(x)$, with $-p^2=m^2$ being the
4D KK mass of the fluctuation. So, with this field decomposition Eq.~\eqref{scalar_pert_2d} can be written as,
\begin{align}
-\ddot{\bar\psi}(z)+\left[\frac{9}{4}\dot A^2 -\frac{5}{2}\ddot A+\dot A \frac{\ddot\phi}{\dot\phi}+2\left(\frac{\ddot\phi}{\dot\phi}\right)^2-
\frac{\dddot\phi}{\dot\phi}\right]\bar\psi(z)&=m^2\bar\psi(z). \label{scalar_pert_2e}
\end{align}
The properties of this equation has also been explored in the past in the context of stability and dynamics
of radion in \cite{Kobayashi:2001jd,Giovannini:2001fh,Giovannini:2001xg} and \cite{Csaki:2000zn}. To develop some
intuition concerning this equations it is convenient to rewrite it in supersymmetric quantum mechanics form.
For this purpose we introduce an auxiliary function $\alpha(z)$ defined by
\beq
\alpha(z)\equiv \frac{e^{\frac{3}{2}A(z)}\dot\phi(z)}{\dot A(z)}.
\label{alpha}
\eeq
Now we can write the potential of the above equation in the following form,
\beq
U_\psi(z)=\left[\frac{9}{4}\dot A^2 -\frac{5}{2}\ddot A+\dot A \frac{\ddot\phi}{\dot\phi}+2\left(\frac{\ddot\phi}{\dot\phi}\right)^2-
\frac{\dddot\phi}{\dot\phi}\right]=\alpha(z)\partial_z^2\left(\frac{1}{\alpha(z)}\right)
=\omega^2(z)-\dot\omega(z),
\label{scalar_potential}
\eeq
where $\omega(z)\equiv\frac{\dot\alpha(z)}{\alpha(z)}$. Then we can rewrite the Eq.~\eqref{scalar_pert_2e} in a
supersymmetric quantum mechanics form as,
\begin{align}
-\partial_z^2{\bar{\psi}}+\left(\omega^2(z)-\dot\omega(z)\right)\bar\psi&=m^2\bar\psi     \notag \\
{\cal A}^{\dagger}{\cal A}\bar\psi&=m^2\bar\psi,     \label{scalar_pert_QM}
\end{align}
where the operator ${\cal A}^\dagger$ and ${\cal A}$ are defined as,
\begin{align}
{\cal A}^\dagger&=\left(-\partial(z)+\omega(z)\right), \hspace{1cm}
{\cal A}=\left(\partial(z)+\omega(z)\right). \label{A_A_dager}
\end{align}
The above supersymmetric form of the scalar perturbation equation \eqref{scalar_pert_QM} guarantee that there is no
solution for $\bar\psi$ with $m^2<0$, hence the fluctuation $\psi$ can not destabilize the background solution.
The zero-mode for the scalar perturbation $\bar\psi(z)$ can be obtained from \eqref{scalar_pert_QM} as,
\beq
\bar\psi_0(z)=\frac{1}{\alpha(z)}=\frac{\dot A(z)}{e^{\frac{3}{2}A(z)}\dot\phi(z)},
\label{psi_zero}
\eeq
which is plotted as a function of $z$ in Fig.~\ref{fig4}. It is important to note that the zero-mode for the scalar
perturbation $\bar\psi(z)$ is not normalizable. As one can easily see from the above expression, $\bar\psi_0$
diverges when $z\to\infty$ (then $\dot\phi(z)\to 0$ and also $e^{\frac{3}{2}A(z)}\to 0$ for
the case $(i)$ and $(ii)$), this behavior is illustrated in Fig.~\ref{fig4}. In fact, also $\psi_0(x,z)=e^{-\frac{3}{2}A(z)}\dot\phi\hat\psi(x)\bar\psi_0(z)=\hat\psi(x) e^{-3A(z)}\dot{A(z)}$ is not normalizable.

Although our main concern here is to verify the stability, nevertheless it is worth to check the behavior
of the potential at $z\to \pm\infty$ in order to see whether there is a mass gap in the spectrum of scalar modes.
From (\ref{scalar_potential}) one can easily find the explicit form of the potential as a function of $z$
\beq
U_\psi(z)= e^{2A[y(z)]}\left[ -\frac72 A^{\prime\prime} + \frac34 A^{\prime\, 2} +
2 A^\prime \frac{\phi^{\prime\prime}}{\phi^\prime} + 2 \left(\frac{\phi^{\prime\prime}}{\phi^\prime}\right)^2
-  \frac{\phi^{\prime\prime\prime}}{\phi^\prime}\right]_{y=y(z)},
\label{potz}
\eeq
where $y$ as a function of $z$ could be determined from
\beq
\int_{y_0}^{y} e^{-A(y^\prime)}dy^\prime = z(y)-z(y_0).
\label{yofz1}
\eeq
If we limit ourself to the large $y$ region and the integration constant $y_0$ is large enough, we can use the asymptotic behavior of $A=A(y)$ as in \eqref{A_prime_RS} then we find
\beq
y(z) \sim \frac{1}{\kappa}\ln(\kappa z + {\rm const.})
\label{yofz2}
\eeq
where $\kappa=\frac{1}{24M^{3}}\left(\frac{4}{3}\kappa_1^2+\frac{4}{3}\kappa_2^2-W_{0}\right)$.
From \eqref{potz} we find that $\lim_{z \to \pm\infty} U_\psi(z)=0$~\footnote{Of course,  $\lim_{y \to \pm\infty} U_\psi[z(y)]=0$,
as well.}, therefore we conclude that the spectrum is continuous starting
at $m^2=0$.

It is worth to comment on another possible zero mode solution. The theory that we are discussing here is
invariant with respect to
a shift along the extra dimension: $y\to y+\epsilon$, therefore if a given metric $g_{MN}(x,y)$ and a
scalar field $\phi(x,y)$ are solutions
of equations of motion, then so are $g_{MN}(x,y+\epsilon)$ and $\phi(x,y+\epsilon)$.
Expanding them around $\epsilon=0$ one obtains
\bea
g_{MN}(x,y+\epsilon)&=&g_{MN}(x,y) + g_{MN}(x,y)^\prime \epsilon + \cdots \\ \non
\phi(x,y+\epsilon) &=& \phi(x,y) + \phi^\prime(x,y) \epsilon + \cdots,
\label{zeroshift1}
\eea
where ellipsis stand for higher powers in $\epsilon$. Since $g_{MN}(x,y+\epsilon)$ and $\phi(x,y+\epsilon)$
and also $g_{MN}(x,y)$ and $\phi(x,y)$
satisfy the equations of motion, therefore $g_{MN}(x,y)^\prime$ and $\phi^\prime(y)$ satisfy linearized equations of motion.
In our parameterizations of the perturbations, \eqref{h_munu}-\eqref{h_55}, that corresponds to
\beq
\psi(x,y)=-A^\prime(y), \lsp \varphi(x,y)=\phi^\prime(y) \lsp {\rm and} \lsp B=E=\chi=0
\label{zeroshift2}
\eeq
As $\psi$ and $\varphi$ given by \eqref{zeroshift2} correspond to modifications of the field configuration
(that satisfies the equations of motion) along the symmetry directions therefore it is supposed to be a zero mode.
Indeed, it could be verified explicitly that $\psi$ and $\varphi$ given by \eqref{zeroshift2} satisfy linearized Einstein equations
\eqref{spert_munu}-\eqref{spert_55} together with the scalar field equation of motion \eqref{phi_cons_1}.
It should be emphasized that in this case the relation $\partial_{\mu}\partial_{\nu}\left(2\psi-\chi\right)=0$
does not hold by the virtue of $2\psi-\chi=0$, but by the fact that $\psi(x,y)$ is $x$-independent while $\chi=0$.
We will not consider those modes any more since they do not depend on $x$ and therefore can not be localized in 4D.

\subsection{Vector perturbations}
\label{Stability of vector perturbations}
The field equation obtained for the transverse vector mode of the perturbation, after integrating
Eq.~\eqref{vector_p5} w.r.t. $x$-coordinate, is,
\begin{align}
\Box C_{\mu}&=0, \hspace{1cm} C_{\mu}^{\prime}+2A^{\prime}C_{\mu}=0,
\label{vector_00}
\end{align}
where we have set the integration constant to zero by using the fact that perturbations should be localized
in 4D so that they do vanish far away from sources. It is more intuitive to write the vector perturbation in
the conformal coordinates so that the results can be interpreted easily. Therefore, in the conformal frame the
equations of motion for the vector modes of the perturbation take the form,
\begin{align}
\Box C_{\mu}&=0, \hspace{1cm} \dot C_{\mu}+3\dot A C_{\mu}=0.
\label{vector_01}
\end{align}
One can immediately notice from
Eqs. \eqref{vector_01} that the vector modes of perturbations are massless.

Since the Eq.~\eqref{vector_01} is first order in $z$-derivatives so it can not be
put into an elegant Schr\"odiger like form as for the case of tensor and scalar modes. Therefore
to see if these modes are localized or not we have to find canonical normal modes of these
perturbations from the second order perturbation of the action \cite{Giovannini:2001fh}, the result reads:
\begin{align}
\delta^2{\cal S}_V&=\int d^5x \frac{1}{2}\left(\eta^{\mu\nu}\partial_\mu{\tilde C}^\alpha\partial_\nu{\tilde C}_\alpha\right),
\label{vector_action}
\end{align}
where, ${\tilde C}_\mu=e^{\frac{3}{2}A}C_\mu$ corresponds to the canonical normal mode.
From Eq.~\eqref{vector_01}, one finds that $C_\mu(x,z)=\hat{C}(x)e^{-3A}$.
So the canonical normal zero-mode of the vector perturbation can be given as,
\begin{align}
{\tilde C}_\mu&=e^{-\frac{3}{2}A}\hat C_\mu (x), \label{vector_02}
\end{align}
where $\hat C_\mu$ satisfies the equation $\Box \hat C_\mu (x)=0$.
Recall that from the requirement of reproducing the General Relativity at low energies we had
\begin{align}
M^{2}_P&= M^{3}\int dz e^{3A(z)}.
\label{4D_Planck mass_1}
\end{align}
Therefore the canonical normal vector modes can not be localized since the integral
$\int dz e^{-3A(z)}$ must be divergent (as a consequence of the finiteness of the 4D Planck
mass). Hence, the vector modes of the perturbation are not localized and therefore
they do not affect issue of stability.

\section{Conclusions}
\label{Conclusions}
Five dimensional Randall-Sundrum like models offer an elegant and simple solution to the
hierarchy problem. The standard formulation of those models assumes the presence
of infinitesimally thin branes embedded in five dimensional space time. Usually, one of the branes
has a negative tension. This work was motivated by a desire to avoid infinitesimally thin branes
and instead to model them by physical objects, e.g. background profiles of a scalar field.

We have shown that even in the presence of non-minimal scalar-gravity coupling it is not
possible to mimic a negative tension brane. In an attempt to construct a model
that is periodic in the extra dimension we have derived a generalization of the
Gibbons-Kallosh-Linde sum rule that holds also if the scalar field couples
non-canonically to the Ricci scalar. It turned out that even in that case periodicity
forbids any non-trivial scalar field profile along the extra dimension.
Therefore we have focused on non-compact extra dimensions. In order to have a chance to address
the hierarchy problem, the scalar background that we introduce is composed of two kink-like
profiles. This set-up in the brane limit corresponds to a model with two thin branes
both having a positive tension. Various possible cases, depending on the location of
the maximum of the warp function has been considered; the most attractive option
turned out to be the one with the maximum located on the top of one of the thick branes.

Stability of the background solution was discussed in details and was verified in the presence of
the most general perturbations of the metric and the scalar field.

\section*{Acknowledgements}
The authors would like to thank Massimo Giovannini and Steven Gubser for their valuable remarks, and
Jacek Tafel and Jose Wudka for interesting discussions.

This work has been supported in part by the National Science Centre (Poland)
as a research project, decision no DEC-2011/01/B/ST2/00438.
AA acknowledges financial support from the Foundation for Polish Science
International PhD Projects Programme co-financed by the EU European
Regional Development Fund.

\appendix

\section{Conventions}
\label{Conv}

In this paper we used the metric signature as $-++++$ and in our conventions the capital roman indices represent 5D objects,
i.e., $M,N,\cdots=0,1,2,3,5$, whereas, the Greek indices label four-dimensional (4D) objects, i.e.,
$\mu,\nu,\cdots=0,1,2,3$. Some of the most frequently used quantities are summarized in this appendix.
In our conventions the definition of 5D covariant derivatives, acting on contravariant and covariant
vectors are, respectively, $\nabla_{M}V^{N}=\partial_{M}V^{N}+\Gamma^{N}_{MA}V^{A}$ and $\nabla_{M}V_{N}=\partial_{M}V_{N}-\Gamma^{A}_{MN}V_{A}$. The definition of the covariant
derivative of a second rank tensor is,
\begin{align}
\nabla_{A}T^{M}_{N}&=\partial_{A}T^{M}_{N}+\Gamma^{M}_{AB}T^{B}_{N}-\Gamma^{C}_{AN}T^{M}_{C}.  \label{bar_T_MN_a}
\end{align}
The 5D d'Alambertian operator $\nabla^2$ is defined as,
\begin{equation}
\nabla^2 =\nabla_{M}\nabla^{M}=\frac{1}{\sqrt{-g}}\partial_{M}\sqrt{-g}g^{MN}\partial_{N}. \label{abox}
\end{equation}

The metric is perturbed as, $g_{MN}=\bar g_{MN}+h_{MN}$, whereas, the inverse of metric perturbation is $h^{MN}=-\bar{g}^{MA}h_{AB}\bar{g}^{BN}$.
The unperturbed (background) metric satisfies the following ansatz:
\begin{equation}
ds^2=e^{2A(y)}\eta_{\mu\nu}dx^{\mu} dx^{\nu}+dy^2,
\label{bmetric}
\end{equation}
where the warp function $A(y)$ is only a function of the extra-spatial coordinate $y$.

The Christoffel symbols or the affine connections take the following form in terms of the linear
perturbation of the metric,
\begin{align}
\Gamma^{A}_{MN}&=\frac{1}{2}\bar{g}^{AB}\left[\partial_{M}h_{NB}+\partial_{N}h_{MB}-
\partial_{B}h_{MN}-2h_{BC}\bar\Gamma^{C}_{MN}\right], \label{affine_p}
\end{align}
where the barred quantities, i.e., $\bar g$ and $\bar \Gamma$, are unperturbed.
The only non-vanishing components of the unperturbed Christoffel's symbol are,
\begin{align}
\bar\Gamma^{\mu}_{\nu5}&=\bar\Gamma^{\mu}_{5\nu}=A^{\prime}\delta^{\mu}_{\nu},\hspace{1cm} \bar\Gamma^{5}_{\mu\nu}=-A^{\prime}e^{2A}\eta_{\mu\nu}. \label{unper_affine_p}
\end{align}
The non-zero components of the perturbed Christoffel's symbols to the first order in the perturbation are,
\begin{align}
\Gamma^{\rho}_{\mu\nu}&=\frac{1}{2}e^{-2A}\eta^{\rho\sigma}\left[\partial_{\mu}h_{\nu\sigma}+\partial_{\nu}h_{\mu\sigma} -\partial_{\sigma}h_{\mu\nu}+2A^{\prime}e^{2A}h_{\sigma5}\eta_{\mu\nu}\right], \label{affine_p_ijk}\\
\Gamma^{\mu}_{\nu5}&=\frac{1}{2}e^{-2A}\eta^{\mu\sigma}\left[\partial_{\nu}h_{\sigma5}+h^{\prime}_{\nu\sigma} -\partial_{\sigma}h_{\nu5}-2A^{\prime}h_{\nu\sigma}\right], \label{affine_p_ij5}\\
\Gamma^{5}_{\mu\nu}&=\frac{1}{2}\left[\partial_{\mu}h_{\nu5}+\partial_{\nu}h_{\mu5}-h^{\prime}_{\mu\nu}
+2A^{\prime}e^{2A}h_{55}\eta_{\mu\nu}\right], \label{affine_p_5ij}\\
\Gamma^{\mu}_{55}&=\frac{1}{2}e^{-2A}\eta^{\mu\nu}\left[2h^{\prime}_{\nu5}-\partial_{\nu}h_{55}\right], \label{affine_p_i55}\\
\Gamma^{5}_{\mu5}&=\frac{1}{2}\left[\partial_{\mu}h_{55}-2A^{\prime}h_{\mu5}\right], \label{affine_p_5i5}\\
\Gamma^{5}_{55}&=\frac{1}{2}h^{\prime}_{55}. \label{affine_p_555}
\end{align}
The following relation for Christoffel's symbols proved to be very useful:
\begin{align}
\Gamma^{M}_{MN}&=\frac{1}{2}\partial_{N}\left[e^{-2A}h^{\mu}_{\mu}+h^{5}_{5}\right]. \label{affine_p_MMN}
\end{align}
The explicit form of 5D d'Alambertian operator acting upon an unperturbed $(\bar\phi(y))$ and a perturbed quantity, say $\phi(x,y)=\bar\phi(y)+\varphi(x,y)$ are,
\begin{align}
\nabla^2\bar\phi(y) =&\bar\phi^{\prime\prime}(y)+4A^\prime(y)\bar\phi^\prime(y), \label{unper_dAlam}\\
\nabla^2\phi(x,y) =&e^{-2A}\Box\varphi+\varphi^{\prime\prime}+4A^{\prime}\varphi^{\prime}
+\frac{1}{2}\bar\phi^{\prime}\left(e^{-2A}\eta^{\mu\nu}h_{\mu\nu}\right)^{\prime}-\frac{1}{2}\bar\phi^{\prime}h^{5\prime}_{5} -\left(\bar\phi^{\prime\prime}+4A^{\prime}\bar\phi^{\prime}\right)h_5^5 \notag\\
&-\frac{1}{2}\bar\phi^{\prime}\eta^{\mu\nu}\left(\partial_\mu h_{\nu5}+\partial_\nu h_{\mu5}\right), \label{per_dAlam}
\end{align}
where $\Box=\eta^{\mu\nu}\partial_\mu\partial_\nu$ is the 4D d'Alambertian operator. The Ricci tensor in the first order in the perturbation can be written as,
\begin{align}
R^{(1)}_{MN}&=\partial_{A}\Gamma^{A}_{MN}-\partial_{M}\Gamma^{A}_{AN}+\Gamma^{A}_{AB}\bar\Gamma^{B}_{MN}+\bar\Gamma^{A}_{AB}\Gamma^{B}_{MN} -\Gamma^{A}_{MB}\bar\Gamma^{B}_{AN}-\bar\Gamma^{A}_{MB}\Gamma^{B}_{AN}. \label{ricci_MN}
\end{align}
%
%
\section{SVT decomposition of perturbations and gauge choice}
\label{SVT decomposition of perturbations and gauge choice}
In this appendix we review the decomposition of most general symmetric perturbation $h_{MN}$ into
\emph{scalar}, \emph{vector} and \emph{tensor} (SVT) modes. The matter of gauge choice in the warped extra-dimension in the presence of a scalar field is also discussed. These issues were studied in the literature, see for example, \cite{Kobayashi:2001jd,Kakushadze:2000zp,DeWolfe:2000xi,Giovannini:2001fh,Giovannini:2001xg,Riazuelo:2002mi}.

Due to the symmetries (4D Poinc\'are invariance) of the background metric and the energy-momentum tensor, we can decompose the perturbations $h_{MN}$ into scalars, vectors and tensors as follows,
\begin{align}
h_{\mu\nu}&=e^{2A}\left[-2\psi \eta_{\mu\nu}-2\partial_{\mu}\partial_{\nu}E +\partial_{\mu}G_{\nu}+\partial_{\nu}G_{\mu}+ H_{\mu\nu}\right],
\label{h_munu}\\
h_{\mu5}&=\partial_{\mu}B +C_{\mu},
\label{h_mu5}\\
h_{55}&=2\chi,
\label{h_55}
\end{align}
where $\psi$, $\chi$, $B$ and $E$ are scalars, whereas, $C_{\mu}$ and $G_{\mu}$ are divergenceless vectors and $H_{\mu\nu}$ is the transverse and traceless tensor, i.e.,
\begin{align}
\partial^\mu C_{\mu}=\partial^\mu G_{\mu}=0, \hspace{1cm} \partial^\mu H_{\mu\nu}=H^{\mu}_{\mu}=0.
\end{align}
The perturbation modes are functions of $x$ and $y$ coordinates.

Let us discuss the uniqueness of the above decomposition.
It is easy to see that $B$ is determined by $h_{\mu5}$ as follows
\beq
\Box B =  \partial^\mu h_{\mu5}.
\label{Beq}
\eeq
Therefore shifting $B$ by a solution the homogeneous equation $\Box \lambda =0$ leads to another allowed solution of
(\ref{Beq})~\footnote{Another way of seeing
the same freedom in determining $B$ and $C_\mu$ is to notice that a shift $B\to B + \lambda$ can be compensated by
an appropriate change of $C_\mu$, $C_\mu \to C_\mu - \partial_\mu \lambda$. Requiring $\Box \lambda=0$, guaranties
that $C_\mu$ remains divergenceless.}. In order to specify the solution of $\Box \lambda=0$
one has to fix initial conditions, that can be done e.g. by specifying $\lambda(t,\vec{x},y)$ and
$\partial_t\lambda(t,\vec{x},y)$
at a given time. Hereafter we are going to assume that at a certain time $t=t_0$ that is far enough
in the past both
$\lambda(t,\vec{x},y)=0$ and $\partial_t\lambda(t,\vec{x},y)=0$. That assumption is physically well
motivated as there is no reason
to observe any perturbations at the very beginning and implies that the only solution of $\Box \lambda=0$
is in fact $\lambda=0$.
Therefore the decomposition (\ref{h_mu5}) is unique.
Similar strategy could be adopted to show uniqueness of the decomposition of $h_{\mu\nu}$ provided
appropriate initial conditions are adopted.
We start by determining $E$ as a solution of the following equation that is implied by (\ref{h_munu}):
\beq
\Box^2 E = \frac13 e^{-2A}\left( \frac14\Box h^\mu_\mu - \partial^\mu\partial^\nu h_{\mu\nu}\right).
\label{Edet}
\eeq
Having $E$ determined (with appropriate initial conditions that ensures uniqueness) one can find $\psi$ solving
\beq
\psi = -\frac18 e^{-2A} h^\mu_\mu -\frac14 \Box E.
\label{psidet}
\eeq
Then $G_\mu$ is a solution of
\beq
\Box G_\mu = e^{-2A} \partial^\nu h_{\mu\nu} + 2\partial_\mu (\psi + \Box E).
\label{Gdet}
\eeq

Now we can write down the first order Einstein equations in terms of the scalar, vector and tensor (SVT) components
defined in \eqref{h_munu}-\eqref{h_55} as,
\begin{align}
(\mu\nu):\hspace{0.5cm} &e^{2A}\eta_{\mu\nu}\left[2\left(A^{\prime\prime}+4A^{\prime2}\right)\chi+A^{\prime}\chi^{\prime} +2\left(A^{\prime\prime}+4A^{\prime2}+\frac{1}{2}e^{-2A}\Box\right)\psi+8A^{\prime}\psi^{\prime}+\psi^{\prime\prime}\right]\notag\\
&+\partial_{\mu}\partial_{\nu}B^{\prime}+2A^\prime \partial_\mu\partial_\nu B+A^\prime \eta_{\mu\nu}\Box B +\frac{1}{2}\left( \partial_\mu C^\prime_\nu+\partial_\nu C^\prime_\mu\right)+A^\prime\left(\partial_\mu C_\nu+\partial_\nu C_\mu\right)\notag\\
&+e^{2A}\left[2\left(A^{\prime\prime}+4A^{\prime2}+\Box\right)\partial_\mu\partial_\nu E+\eta_{\mu\nu}A^\prime\Box E^\prime+4A^{\prime}\partial_\mu\partial_\nu E^{\prime}+\partial_\mu\partial_\nu E^{\prime\prime}\right]\notag\\
&-\left(A^{\prime\prime}+4A^{\prime2}+\frac{1}{2}\Box\right)\left( \partial_\mu G_\nu+\partial_\nu G_\mu\right)-\frac{1}{2}e^{2A}\left[ \partial_\mu G^{\prime\prime}_\nu+\partial_\nu G^{\prime\prime}_\mu+4A^\prime\left( \partial_\mu G^{\prime}_\nu+\partial_\nu G^{\prime}_\mu\right)\right] \notag\\
&+\partial_{\mu}\partial_{\nu}\left(2\psi-\chi\right)-\left(A^{\prime\prime}+4A^{\prime2}+\frac{1}{2}\Box\right) H_{\mu\nu}-2e^{2A}A^\prime H^\prime_{\mu\nu}-\frac{1}{2}e^{2A}H_{\mu\nu}^{\prime\prime}    \notag\\
&=\frac{1}{4M^{3}}\frac{2}{3}e^{2A}\left[\eta_{\mu\nu}\frac{\partial V(\phi)}{\partial \phi}\varphi+V(\phi)\left(-2\psi \eta_{\mu\nu}-2\partial_{\mu}\partial_{\nu}E +\partial_{\mu}G_{\nu}+\partial_{\nu}G_{\mu}+ H_{\mu\nu}\right)\right],
\label{pert_munu} \\
(\mu5):\hspace{0.5cm} &3\partial_{\mu}\psi^{\prime}+3A^{\prime}\partial_{\mu}\chi-\frac{1}{2}e^{-2A}\Box C_{\mu}+\frac{1}{2}\Box G_\mu^\prime=\frac{1}{4M^{3}}\phi^{\prime}\partial_{\mu}\varphi,
\label{pert_mu5}\\
(55):\hspace{0.5cm} &4\left(\psi^{\prime\prime}+2A^{\prime}\psi^{\prime}\right)+4A^{\prime}\chi^{\prime}-e^{-2A}\Box \chi +\Box\left(E^{\prime\prime}+2A^{\prime}E^{\prime}\right)+e^{-2A}\Box B^{\prime} \notag\\
&\hspace{5cm}=\frac{1}{4M^{3}}\left[2\phi^{\prime}\varphi^\prime+\frac{2}{3}\frac{\partial V(\phi)}{\partial \phi}\varphi +\frac{4}{3}V(\phi)\chi\right].
\label{pert_55}
\end{align}
Adopting the background equations of motion the above equations could be simplified as follows:
\begin{align}
(\mu\nu):\hspace{.5cm} & e^{2A}\eta_{\mu\nu}\bigg[2\left(A^{\prime\prime}+4A^{\prime2}\right)\chi+A^{\prime}\chi^{\prime} +e^{-2A}\Box\psi+8A^{\prime}\psi^{\prime}+\psi^{\prime\prime}+e^{-2A}A^\prime \Box B+A^\prime\Box E^\prime\bigg]\notag\\
&+\partial_{\mu}\partial_{\nu}\bigg[2\psi-\chi+B^{\prime}+2A^\prime B+e^{2A}\left(2\Box E+4A^{\prime} E^{\prime}+ E^{\prime\prime}\right)\bigg] \notag\\
&+\frac{1}{2}\partial_\mu\bigg[C_{\nu}^{\prime}+2A^{\prime}C_{\nu}-4e^{2A}A^{\prime}G_{\nu}^{\prime}-e^{2A}G_{\nu}^{\prime\prime}\bigg]\notag\\
&+\frac{1}{2}\partial_\nu \bigg[C_{\mu}^{\prime}+2A^{\prime}C_{\mu}-4e^{2A}A^{\prime}G_{\mu}^{\prime}-e^{2A}G_{\mu}^{\prime\prime}\bigg] \notag\\
&-\frac{1}{2}\left(\Box H_{\mu\nu}+4e^{2A}A^\prime H^\prime_{\mu\nu}+e^{2A}H_{\mu\nu}^{\prime\prime}\right)=\frac{1}{4M^{3}}\frac{2}{3}e^{2A}\eta_{\mu\nu}\frac{\partial V(\phi)}{\partial \phi}\varphi,
\label{pert_munu_1} \\
(\mu5):\hspace{.5cm} &3\partial_{\mu}\psi^{\prime}+3A^{\prime}\partial_{\mu}\chi-\frac{1}{2}e^{-2A}\Box C_{\mu}+\frac{1}{2}\Box G_\mu^\prime=\frac{1}{4M^{3}}\phi^{\prime}\partial_{\mu}\varphi,
\label{pert_mu5_1} \\
(55):\hspace{.5cm} &4\left(\psi^{\prime\prime}+2A^{\prime}\psi^{\prime}\right)+4A^{\prime}\chi^{\prime}-e^{-2A}\Box \chi +\Box\left(E^{\prime\prime}+2A^{\prime}E^{\prime}\right)+e^{-2A}\Box B^{\prime} \notag\\
&\hspace{5cm}=\frac{1}{4M^{3}}\left[2\phi^{\prime}\varphi^\prime+\frac{2}{3}\frac{\partial V(\phi)}{\partial \phi}\varphi +\frac{4}{3}V(\phi)\chi\right].
\label{pert_55_1}
\end{align}
Now comparing the coefficients of $\eta_{\mu\nu}$, $\partial_{\mu}\partial_{\nu}$, $\partial_{\nu}$ and
the tensors on both sides we get from $(\mu\nu)$ components
the following equations of motion for the scalar, vector and tensor modes
of the perturbations
\begin{align}
2\left(A^{\prime\prime}+4A^{\prime2}\right)\chi+A^{\prime}\chi^{\prime} +e^{-2A}\Box\psi+8A^{\prime}\psi^{\prime}+\psi^{\prime\prime}+e^{-2A}A^\prime \Box B+A^\prime\Box E^\prime &=\frac{1}{4M^{3}}\frac{2}{3}\frac{\partial V(\phi)}{\partial \phi}\varphi,
\label{pert_munu_s}\\
\partial_{\mu}\partial_{\nu}\bigg[2\psi-\chi+B^{\prime}+2A^\prime B+e^{2A}\left(2\Box E+4A^{\prime} E^{\prime}+ E^{\prime\prime}\right)\bigg]&=0, \label{pert_munu_s2}\\
\partial_{\nu}\bigg[C_{\mu}^{\prime}+2A^{\prime}C_{\mu}-4e^{2A}A^{\prime}G_{\mu}^{\prime}-e^{2A}G_{\mu}^{\prime\prime}\bigg]&=0, \label{pert_munu_v}\\
-\frac{1}{2}\left(\Box H_{\mu\nu}+4e^{2A}A^\prime H^\prime_{\mu\nu}+e^{2A}H_{\mu\nu}^{\prime\prime}\right)&=0.
\label{pert_munu_t}
\end{align}
For $(\mu 5)$ and $(55)$ we obtain the following equations:
\begin{align}
\partial_{\mu}\left(3\psi^{\prime}+3A^{\prime}\chi-\frac{1}{4M^{3}}\phi^{\prime}\varphi\right)&=0,
\label{pert_mu5_s} \\
e^{-2A}\Box C_{\mu}-\Box G_\mu^\prime&=0,
\label{pert_mu5_v} \\
4\left(\psi^{\prime\prime}+2A^{\prime}\psi^{\prime}\right)+4A^{\prime}\chi^{\prime}-e^{-2A}\Box \chi & +\Box\left(E^{\prime\prime}+2A^{\prime}E^{\prime}\right)+e^{-2A}\Box B^{\prime}\notag\\
&=\frac{1}{4M^{3}}\left[2\phi^{\prime}\varphi^\prime+\frac{2}{3}\frac{\partial V(\phi)}{\partial \phi}\varphi +\frac{4}{3}V(\phi)\chi\right].
\label{pert_55_s}
\end{align}
The above equations of motion for the scalar, vector and tensor modes of perturbations are applicable for any gauge choice, in the
main text we decide to choose the longitudinal gauge defined by the condition $B=E=G_\mu=0$ as discussed below.

Now we will consider the coordinate/gauge transformations and then we will turn to the question of choosing the appropriate gauge
in order to eliminate artifacts of the freedom of choosing a reference frame. Lets consider the following coordinate transformation,
\begin{align}
\check{x}^M&=x^{M}-\xi^{M},
\label{X_check}
\end{align}
where the $\xi^{M}$ is an infinitesimally small function of space time, i.e., $|\xi^{M}|<< |x^M|$ and $\xi^{M}=(\xi^{\mu},\xi^{5})$ with $\xi^{\mu}$ being a 4D vector and $\xi^{5}$ a scalar change in the 5th coordinate $y$.
In order to write down corresponding gauge transformations of the  decomposed scalars, vector and tensor modes, it is useful to decompose also the 4D vector $\xi^{\mu}$ into the divergenceless vector $\xi_{\perp}^{\mu}$ and gradient of the scalar $\xi_{\parallel}$, i.e.,
\begin{align}
\xi^{\mu}&=\xi_{\perp}^{\mu}+\partial_{\mu}\xi_{\parallel}, \hspace{1cm} \partial_{\mu}\xi^{\mu}_{\perp}=0. \label{x_mu_check}
\end{align}
It is easy to show that the change in metric perturbation $h_{MN}$ corresponding to (\ref{X_check}) reads
\begin{align}
\check{h}_{MN}&=h_{MN}+\delta h_{MN},
\end{align}
with
\begin{align}
\delta h_{MN}&=\nabla_{M}\xi_{N}+\nabla_{N}\xi_{M}, \label{delta_h}
\end{align}
where as usual $\nabla_{M}$ is the 5D covariant derivative, see the Appendix \ref{Conv}. The explicit form of the components of $\delta h_{MN}$ are given by,
\begin{align}
\delta h_{\mu\nu}&=\partial_{\mu}\xi_{\perp\nu}+\partial_{\nu}\xi_{\perp\mu}+2\partial_{\mu}\partial_{\nu}\xi_{\parallel} +2A^{\prime}e^{2A}\eta_{\mu\nu}\xi_{5}, \label{del_h_munu}\\
\delta h_{\mu5}&=\partial_{\mu}\xi_{5}+\partial_{\mu}\xi_\parallel^{\prime}+\xi_{\perp\mu}^{\prime}-2A^{\prime}\xi_{\perp\mu}-2A^{\prime}\partial_{\mu}\xi_\parallel, \label{del_h_mu5}\\
\delta h_{55}&=2\xi_{5}^{\prime}, \label{del_h_55}
\end{align}
The above transformations of the metric perturbation $h_{MN}$ induce the corresponding transformations of the metric perturbation components defined by Eqs. \eqref{h_munu}-\eqref{h_55} as,
\begin{align}
\check \psi&=\psi-A^{\prime}\xi^{5}, \hspace{2.5cm} \check E=E-e^{-2A}\xi_{\parallel}, \label{scalar_gauge_1}\\
\check \chi&=\chi+\xi_{5}^{\prime}, \hspace{3cm} \check B=B+\xi_\parallel^{\prime}+\xi^{5}-2A^{\prime}\xi_{\parallel}, \label{scalar_gauge_2}\\
\check C_{\mu}&=C_\mu+\xi_{\perp\mu}^{\prime}-2A^{\prime}\xi_{\perp\mu}, \hspace{0.7cm} \check G_{\mu}=G_\mu+e^{-2A}\xi_{\perp\mu}, \label{vector_gauge}
\end{align}
whereas, $H_{\mu\nu}$ is unaffected by the coordinate transformations.
Similarly, the gauge transformation of the scalar field perturbation $\varphi$ can be easily obtained as,
\begin{align}
\check\varphi&=\varphi+\delta \varphi=\varphi+\phi^\prime\xi^5.
\label{scal_per}
\end{align}
Similarly, the gauge transformations of the energy momentum tensor can be written as,
\begin{align}
\check{\tilde{T}}_{MN}^{(1)}&={\tilde{T}}_{MN}^{(1)}+\delta{\tilde{T}}_{MN}^{(1)},
\end{align}
where,
\begin{align}
\delta\tilde{T}_{MN}^{(1)}&= \tilde{T}_{MA}^{(0)}\nabla_{N}\xi^{A}+\tilde{T}_{NB}^{(0)}\nabla_{M}\xi^{B}+\nabla_{C}\tilde{T}_{MN}^{(0)}\xi^{C}. \label{delta_T}
\end{align}

Now we turn our attention towards the issue of choosing a gauge. It proves to be convenient to adopt the so-called
longitudinal or Newtonian gauge defined by the conditions: $\check B=\check E =0$ in the scalar and $\check G_\mu=0$
in the vector sector. It is important to note that, indeed, one can always choose the gauge parameters such that the
gauge conditions are satisfied, i.e.,
\begin{align}
\xi_{\parallel}(x,y)&=e^{2A}E,
\label{gauge_fix_1}\\
\xi^{5}(x,y)&=-B-e^{2A}\left(e^{-2A}E\right)^{\prime},
\label{gauge_fix_2}\\
\xi_{\perp\mu}(x,y)&=-e^{2A}G_\mu.
\label{gauge_fix_3}
\end{align}
Here one can note that this choice of gauge fixing completely fixes the gauge and so that there is no residual
gauge freedom left.


\begin{thebibliography}{99}

\bibitem{ArkaniHamed:1998rs}
  N.~Arkani-Hamed, S.~Dimopoulos and G.~R.~Dvali,
  ``The Hierarchy problem and new dimensions at a millimeter,''
  Phys.\ Lett.\ B {\bf 429}, 263 (1998)
  [hep-ph/9803315].

\bibitem{Antoniadis:1998ig}
  I.~Antoniadis, N.~Arkani-Hamed, S.~Dimopoulos and G.~R.~Dvali,
  ``New dimensions at a millimeter to a Fermi and superstrings at a TeV,''
  Phys.\ Lett.\ B {\bf 436}, 257 (1998)
  [hep-ph/9804398].

\bibitem{Randall:1999ee}
  L.~Randall and R.~Sundrum,
  ``A large mass hierarchy from a small extra dimension,''
  Phys.\ Rev.\ Lett.\  {\bf 83}, 3370 (1999)
  [hep-ph/9905221].

\bibitem{Rubakov:1983bb}
  V.~A.~Rubakov and M.~E.~Shaposhnikov,
  ``Do We Live Inside a Domain Wall?,''
  Phys.\ Lett.\ B {\bf 125}, 136 (1983), ``Extra Space-Time Dimensions: Towards a Solution to the Cosmological Constant Problem,''
  Phys.\ Lett.\ B {\bf 125}, 139 (1983).

\bibitem{DeWolfe:1999cp}
  O.~DeWolfe, D.~Z.~Freedman, S.~S.~Gubser and A.~Karch,
  ``Modeling the fifth-dimension with scalars and gravity,''
  Phys.\ Rev.\ D {\bf 62}, 046008 (2000)
  [hep-th/9909134].

\bibitem{Gremm:1999pj}
  M.~Gremm,
  ``Four-dimensional gravity on a thick domain wall,''
  Phys.\ Lett.\ B {\bf 478}, 434 (2000)
  [hep-th/9912060], ``Thick domain walls and singular spaces,''
  Phys.\ Rev.\ D {\bf 62}, 044017 (2000)
  [hep-th/0002040].

\bibitem{Csaki:2000fc}
  C.~Csaki, J.~Erlich, T.~J.~Hollowood and Y.~Shirman,
  ``Universal aspects of gravity localized on thick branes,''
  Nucl.\ Phys.\ B {\bf 581}, 309 (2000)
  [hep-th/0001033].

\bibitem{Kehagias:2000au}
  A.~Kehagias and K.~Tamvakis,
  ``Localized gravitons, gauge bosons and chiral fermions in smooth spaces generated by a bounce,''
  Phys.\ Lett.\ B {\bf 504}, 38 (2001)
  [hep-th/0010112].

\bibitem{Kobayashi:2001jd}
  S.~Kobayashi, K.~Koyama and J.~Soda,
  ``Thick brane worlds and their stability,''
  Phys.\ Rev.\ D {\bf 65}, 064014 (2002)
  [hep-th/0107025].
  
\bibitem{Melfo:2002wd}
  A.~Melfo, N.~Pantoja and A.~Skirzewski,
  ``Thick domain wall space-times with and without reflection symmetry,''
  Phys.\ Rev.\ D {\bf 67}, 105003 (2003)
  [gr-qc/0211081].

\bibitem{Bronnikov:2003gg}
  K.~A.~Bronnikov and B.~E.~Meierovich,
  ``A General thick brane supported by a scalar field,''
  Grav.\ Cosmol.\  {\bf 9}, 313 (2003)
  [gr-qc/0402030].

\bibitem{Bazeia:2008zx}
  D.~Bazeia, A.~R.~Gomes, L.~Losano and R.~Menezes,
  ``Braneworld Models of Scalar Fields with Generalized Dynamics,''
  Phys.\ Lett.\ B {\bf 671}, 402 (2009)
  [arXiv:0808.1815 [hep-th]].

\bibitem{Bazeia:2007nd}
  D.~Bazeia, A.~R.~Gomes and L.~Losano,
  ``Gravity localization on thick branes: A Numerical approach,''
  Int.\ J.\ Mod.\ Phys.\ A {\bf 24}, 1135 (2009)
  [arXiv:0708.3530 [hep-th]].

\bibitem{BarbosaCendejas:2007hs}
  N.~Barbosa-Cendejas, A.~Herrera-Aguilar, U.~Nucamendi and I.~Quiros,
  ``Mass hierarchy and mass gap on thick branes with Poincare symmetry,''
  arXiv:0712.3098 [hep-th].

\bibitem{Dzhunushaliev:2009va}
 \emph{ For review on thick branes see}, V.~Dzhunushaliev, V.~Folomeev and M.~Minamitsuji,
  ``Thick brane solutions,''
  Rept.\ Prog.\ Phys.\  {\bf 73}, 066901 (2010)
  [arXiv:0904.1775 [gr-qc]].

\bibitem{Oda:2008tk}
  K.~-y.~Oda, T.~Suyama and N.~Yokoi,
  ``Smoothing out Negative Tension Brane,''
  Phys.\ Lett.\ B {\bf 675}, 455 (2009)
  [arXiv:0811.3459 [hep-ph]].

\bibitem{Gibbons:2000tf}
  G.~W.~Gibbons, R.~Kallosh and A.~D.~Linde,
  ``Brane world sum rules,''
  JHEP {\bf 0101}, 022 (2001)
  [hep-th/0011225].

\bibitem{Randall:1999vf}
  L.~Randall and R.~Sundrum,
  ``An alternative to compactification,''
  Phys.\ Rev.\ Lett.\  {\bf 83}, 4690 (1999)
  [hep-th/9906064].

\bibitem{Abdalla:2010sz}
  M.~C.~B.~Abdalla, M.~E.~X.~Guimaraes and J.~M.~Hoff da Silva,
  ``Positive tension 3-branes in an $AdS_{5}$ bulk,''
  JHEP {\bf 1009}, 051 (2010)
  [arXiv:1001.1075 [hep-th]].

\bibitem{Wudka:2011we}
  J.~Wudka,
  ``Periodic thick-brane configurations and their stability,''
  J.\ Phys.\ G G {\bf 38}, 075003 (2011)
  [arXiv:1107.0944 [hep-th]].

\bibitem{Lykken:1999nb}
  J.~D.~Lykken and L.~Randall,
  ``The Shape of gravity,''
  JHEP {\bf 0006}, 014 (2000)
  [hep-th/9908076].

\bibitem{Gregory:2000jc}
  R.~Gregory, V.~A.~Rubakov and S.~M.~Sibiryakov,
  ``Opening up extra dimensions at ultra large scales,''
  Phys.\ Rev.\ Lett.\  {\bf 84}, 5928 (2000)
  [hep-th/0002072].

\bibitem{Garriga:1999yh}
  J.~Garriga and T.~Tanaka,
  ``Gravity in the brane world,''
  Phys.\ Rev.\ Lett.\  {\bf 84}, 2778 (2000)
  [hep-th/9911055].

\bibitem{Giddings:2000mu}
  S.~B.~Giddings, E.~Katz and L.~Randall,
  ``Linearized gravity in brane backgrounds,''
  JHEP {\bf 0003}, 023 (2000)
  [hep-th/0002091].

\bibitem{Karch:2000ct}
  A.~Karch and L.~Randall,
  ``Locally localized gravity,''
  JHEP {\bf 0105}, 008 (2001)
  [hep-th/0011156].

\bibitem{Csaki:2000ei}
  C.~Csaki, J.~Erlich and T.~J.~Hollowood,
  ``Graviton propagators, brane bending and bending of light in theories with quasilocalized gravity,''
  Phys.\ Lett.\ B {\bf 481}, 107 (2000)
  [hep-th/0003020].

\bibitem{DeWolfe:2000xi}
  O.~DeWolfe and D.~Z.~Freedman,
  ``Notes on fluctuations and correlation functions in holographic renormalization group flows,''
  hep-th/0002226.

\bibitem{Giovannini:2001fh}
  M.~Giovannini,
  ``Gauge invariant fluctuations of scalar branes,''
  Phys.\ Rev.\ D {\bf 64}, 064023 (2001)
  [hep-th/0106041], ``Scalar normal modes of higher dimensional gravitating kinks,''
  Class.\ Quant.\ Grav.\  {\bf 20}, 1063 (2003)
  [gr-qc/0207116].

\bibitem{Giovannini:2001xg}
  M.~Giovannini,
  ``Localization of metric fluctuations on scalar branes,''
  Phys.\ Rev.\ D {\bf 65}, 064008 (2002)
  [hep-th/0106131], ``Theory of gravitational fluctuations in brane world models,
  Int.\ J.\ Mod.\ Phys.\ D {\bf 11}, 1209 (2002)..

\bibitem{Csaki:2000zn}
  C.~Csaki, M.~L.~Graesser and G.~D.~Kribs,
  ``Radion dynamics and electroweak physics,''
  Phys.\ Rev.\ D {\bf 63}, 065002 (2001)
  [hep-th/0008151].

\bibitem{Cvetic:2008gu}
  M.~Cvetic and M.~Robnik,
  ``Gravity Trapping on a Finite Thickness Domain Wall: An Analytic Study,''
  Phys.\ Rev.\ D {\bf 77}, 124003 (2008)
  [arXiv:0801.0801 [hep-th]].

\bibitem{Kakushadze:2000zp}
  Z.~Kakushadze,
  ``Localized (super)gravity and cosmological constant,''
  Nucl.\ Phys.\ B {\bf 589}, 75 (2000)
  [hep-th/0005217].

\bibitem{Brandhuber:1999hb}
  A.~Brandhuber and K.~Sfetsos,
  ``Nonstandard compactifications with mass gaps and Newton's law,''
  JHEP {\bf 9910}, 013 (1999)
  [hep-th/9908116].

\bibitem{Aybat:2010sn}
  S.~M.~Aybat and D.~P.~George,
  ``Stability of Scalar Fields in Warped Extra Dimensions,''
  JHEP {\bf 1009}, 010 (2010)
  [arXiv:1006.2827 [hep-th]].

\bibitem{Andrianov:2012ae}
  A.~A.~Andrianov, V.~A.~Andrianov and O.~O.~Novikov,
  ``Localization of scalar fields on self-gravitating thick branes,''
  arXiv:1210.3698 [hep-th].

\bibitem{Davoudiasl:1999tf}
  H.~Davoudiasl, J.~L.~Hewett and T.~G.~Rizzo,
  ``Bulk gauge fields in the Randall-Sundrum model,''
  Phys.\ Lett.\ B {\bf 473}, 43 (2000)
  [hep-ph/9911262].

\bibitem{Pomarol:1999ad}
  A.~Pomarol,
  ``Gauge bosons in a five-dimensional theory with localized gravity,''
  Phys.\ Lett.\ B {\bf 486}, 153 (2000)
  [hep-ph/9911294].

\bibitem{Grossman:1999ra}
  Y.~Grossman and M.~Neubert,
  ``Neutrino masses and mixings in nonfactorizable geometry,''
  Phys.\ Lett.\ B {\bf 474}, 361 (2000)
  [hep-ph/9912408].

\bibitem{Chang:1999nh}
  S.~Chang, J.~Hisano, H.~Nakano, N.~Okada and M.~Yamaguchi,
  ``Bulk standard model in the Randall-Sundrum background,''
  Phys.\ Rev.\ D {\bf 62}, 084025 (2000)
  [hep-ph/9912498].

\bibitem{Bogdanos:2006qw}
  C.~Bogdanos, A.~Dimitriadis and K.~Tamvakis,
  ``Brane models with a Ricci-coupled scalar field,''
  Phys.\ Rev.\ D {\bf 74}, 045003 (2006)
  [hep-th/0604182].

\bibitem{Guo:2011wr}
  H.~Guo, Y.~-X.~Liu, Z.~-H.~Zhao and F.~-W.~Chen,
  ``Thick branes with a non-minimally coupled bulk-scalar field,''
  Phys.\ Rev.\ D {\bf 85}, 124033 (2012)
  [arXiv:1106.5216 [hep-th]].

\bibitem{Bajc:1999mh}
  B.~Bajc and G.~Gabadadze,
  ``Localization of matter and cosmological constant on a brane in anti-de Sitter space,''
  Phys.\ Lett.\ B {\bf 474}, 282 (2000)
  [hep-th/9912232].

\bibitem{Liu:2011zy}
  Y.~-X.~Liu, H.~Guo, C.~-EFu and H.~-T.~Li,
  ``Localization of Bulk Matters on a Thick Anti-de Sitter Brane,''
  Phys.\ Rev.\ D {\bf 84}, 044033 (2011)
  [arXiv:1101.4145 [hep-th]].

\bibitem{Liang:2009zzc}
  J.~Liang and Y.~-S.~Duan,
  ``Localization of matters on thick branes,''
  Phys.\ Lett.\ B {\bf 678}, 491 (2009).

\bibitem{Melfo:2006hh}
  A.~Melfo, N.~Pantoja and J.~D.~Tempo,
  ``Fermion localization on thick branes,''
  Phys.\ Rev.\ D {\bf 73}, 044033 (2006)
  [hep-th/0601161].

\bibitem{Koley:2004at}
  R.~Koley and S.~Kar,
  ``Scalar kinks and fermion localisation in warped spacetimes,''
  Class.\ Quant.\ Grav.\  {\bf 22}, 753 (2005)
  [hep-th/0407158].

\bibitem{Riazuelo:2002mi}
  A.~Riazuelo, F.~Vernizzi, D.~A.~Steer and R.~Durrer,
  ``Gauge invariant cosmological perturbation theory for brane worlds,''
  [hep-th/0205220].

\bibitem{Csaki:2000pp}
  C.~Csaki, J.~Erlich and T.~J.~Hollowood,
  ``Quasilocalization of gravity by resonant modes,''
  Phys.\ Rev.\ Lett.\  {\bf 84}, 5932 (2000)
  [hep-th/0002161].

\bibitem{Gregory:2000iu}
  R.~Gregory, V.~A.~Rubakov and S.~M.~Sibiryakov,
  ``Gravity and antigravity in a brane world with metastable gravitons,''
  Phys.\ Lett.\ B {\bf 489}, 203 (2000)
  [hep-th/0003045].

\bibitem{Dvali:2000rv}
  G.~R.~Dvali, G.~Gabadadze and M.~Porrati,
  ``Metastable gravitons and infinite volume extra dimensions,''
  Phys.\ Lett.\ B {\bf 484}, 112 (2000)
  [hep-th/0002190], ``4-D gravity on a brane in 5-D Minkowski space,''
  Phys.\ Lett.\ B {\bf 485}, 208 (2000)
  [hep-th/0005016].


\end{thebibliography}
\end{document}